\documentclass[12pt]{article}
\usepackage{fullpage,epsfig,graphics,amsbsy,amssymb,cancel,slashed,mathrsfs}
\usepackage{psfrag,hyperref}
\usepackage{graphicx,color}
\newcommand{\be}{\begin{eqnarray}}
\newcommand{\ee}{\end{eqnarray}}
\usepackage{amsmath}
\numberwithin{equation}{section}

\newcommand{\BB}{\mathbb}

\newcommand{\FR}{\mathfrak}

\def\L{{\cal L}}

\def\cA{\mathscr{A}}

\newcommand{\bea}{\begin{eqnarray}}
\newcommand{\eea}{\end{eqnarray}}
\newcommand{\nn}{\nonumber}
\newcommand{\Tr}{\textrm{Tr}}

\newcommand{\sbullet}{\textrm{\tiny{\textbullet}}}

\newcommand{\bra}{\langle}
\newcommand{\ket}{\rangle}

\def\ga{\alpha}

\def\Gc{\Gamma}
\def\gd{\delta}
\def\gt{\theta}
\def\Gt{\Theta}
\def\gs{\sigma}

\def\gl{\lambda}
\def\Go{\Omega}

\def\spinc{Spin${}^c$}

\begin{document}

\thispagestyle{empty}
\begin{flushright} \small
UUITP-17/12
 \end{flushright}
\smallskip
\begin{center} \LARGE
{\bf The perturbative partition function of \\  supersymmetric
 5D Yang-Mills theory with matter  on the five-sphere}
 \\[12mm] \normalsize
{\bf  Johan~K\"all\'en$^a$, Jian Qiu$^b$ and Maxim Zabzine$^a$} \\[8mm]
 {\small\it
  ${}^a$Department of Physics and Astronomy,
     Uppsala university,\\
     Box 516,
     SE-75120 Uppsala,
     Sweden\\
     \vspace{.5cm}
${}^b$I.N.F.N. and Dipartimento di Fisica\\
     Via G. Sansone 1, 50019 Sesto Fiorentino - Firenze, Italy\\
   }
\end{center}
\vspace{7mm}
\begin{abstract}
 \noindent
  Based on the construction by Hosomichi, Seong and Terashima  we consider $N=1$ supersymmetric 5D Yang-Mills theory with matter on a five-sphere with radius $r$. This theory can be thought of as a deformation of the theory in flat space with deformation parameter $r$ and this deformation preserves 8 supercharges. We calculate the full perturbative partition function as a function of $r/g_{YM}^2$, where $g_{YM}$ is the Yang-Mills coupling, and the answer is given in terms of a matrix model. We perform the calculation using localization techniques.
      We also argue that in the large $N$-limit of this deformed  5D Yang-Mills theory this matrix model
      provides the leading contribution to the partition function and the rest is exponentially suppressed.
  \end{abstract}

\eject
\normalsize

\section{Introduction}

Recently 5D supersymmetric Yang-Mills theory has attracted a lot of attention, in particular  its relation to the
6D (2,0) superconformal theory \cite{Witten:1995zh}. The 5D theory is not perturbatively renormalizable,
 however the works  \cite{Lambert:2010iw,Lambert:2011eg, Douglas:2010iu}  suggests arguments in favor of UV-finiteness
 of 5D supersymmetric Yang-Mills theory. Roughly these arguments appeal to the fact that the 5D theory comes
  from a reduction of a well-defined 6D theory.  Moreover it has been argued that the 5D theory may contain all
   the degrees of freedom of the 6D theory. In this work we calculate the full perturbative
   partition function of deformations of $N=1$ and $N=2$ 5D supersymmetric Yang-Mills theory. Our calculation suggests
    that the perturbative
     partition function is well-defined and thus offering more support to the ideas advocated in  \cite{Lambert:2011eg, Douglas:2010iu}.

 The present work is a natural continuation of the two closely related works \cite{KallenZabzine12} and  \cite{HosomichiSeongTerashima}. Let us briefly outline our logic. In $\mathbb{R}^5$ the $N=1$ supersymmetric Yang-Mills theory is invariant under
  8 supercharges while the $N=2$ theory is invariant under 16 supercharges. Neither one of these theories is superconformal.
Recently in \cite{HosomichiSeongTerashima} a supersymmetric version of a 5D Yang-Mills theory which preserves 8 supercharges has been
 constructed on $S^5$.  This theory on $S^5$ does not have the same status as
  the corresponding superconformal theories in the 4D and 3D cases on $S^4$ and $S^3$, originally considered in  \cite{Pestun:2007rz}  and \cite{Kapustin:2009kz}, respectively.
   However, we may look at the theory on $S^5$ as a
   one parameter deformation of the flat theory  with the parameter $r$ given by the radius of $S^5$. The theory on $S^5$ is perfectly adapted for the localization technique, which has recently been applied to theories in two, three and four dimensions on spheres and other compact manifolds in for example \cite{Pestun:2007rz}-\cite{Doroud:2012xw}. In \cite{HosomichiSeongTerashima} the localization locus for the theory on the five-sphere was determined, however, an important part when it comes to localization calculations, namely the one-loop determinants, was not found.
    In this work we will, among other things, perform the calculation of the one-loop determinants, using techniques from  \cite{KallenZabzine12} (which in turn is very much inspired by the calculation performed by  Pestun  in \cite{Pestun:2007rz}). The main idea is to recast the supersymmetry transformations into a cohomological form, and then calculate the one-loop determinant using the Atiyah-Singer index theorem.

We will calculate the contribution to the partition function arising from fluctuations around the isolated trivial connection, this is what we call the perturbative partition function. The perturbative partition function is a function of the ratio $\frac{r}{g_{YM}^2}$, where $g_{YM}$ is the Yang-Mills coupling,  and the answer
     can be written in terms of a matrix model.  Since this is our main result, we state the answer here. The perturbative partition function of $N=1$ 5D Yang-Mills theory with matter in
      a representation $R$ on the five-sphere is
      given by\footnote{ [This footnote was added in v3] Previous versions of this paper had the wrong coefficient in front of the $\text{Tr}(\phi^2)$-term. In this version we have corrected this mistake. In \cite{Minahan:2013jwa} the flat space limit of this matrix model is compared with one-loop calculations previously performed in flat space, and perfect matching is found.}
     \be
Z&=&\int\limits_{\rm Cartan} [d\phi]~e^{-  \frac{ 8\pi^3 r}{g_{YM}^2}  \text{Tr}(\phi^2)} {\rm det}_{\rm Ad}\left (  \sin ( i\pi  \phi) e^{ \frac{1}{2} f(i \phi )}  \right ) \nn \\
&&\times~  {\rm det}_{R} \left ( \left ( \cos ( i\pi \phi )\right )^{\frac{1}{4}} e^{-\frac{1}{4} f \left (\frac{1}{2} -  i\phi \right ) - \frac{1}{4} f \left (\frac{1}{2} +  i\phi \right )} \right )  + \mathcal{O} (e^{-\frac{16 \pi^3 r}{g_{YM}^2}})~,\label{vector1loop-intro}
\ee
     where $\phi$ is a dimensionless matrix and the function $f$ is defined by
 \be
 f (y)=\frac{i \pi y^3}{3}+y^2 \ln{(1-e^{-2 \pi i y})}+ \frac{iy}{\pi} \text{Li}_2(e^{-2  \pi iy})+\frac{1}{2\pi^2}\text{Li}_3(e^{-2 \pi i y})-
 \frac{\zeta(3)}{2\pi^2}~.\label{definition-f-func-intro}
\ee
      In our answer we cannot simply send  $r$ to infinity
      to recover the flat limit unless we send the coupling $g_{YM}^2$ to infinity as well. This in turn corresponds to
      going to the 6D theory.

   What can be expected from our matrix model? At present, we are unable to calculate the full non-perturbative
    partition function since we have to take into account instantons, which gives the non-perturbative contributions.
     However, if we are interested in the large $N$-limit
     of the 5D supersymmetric Yang-Mills theory with fixed 't Hooft  coupling $\frac{g^2_{YM}}{r} N$
          then we can ignore the terms $e^{-\frac{16 \pi^3 r}{g_{YM}^2}}$ since their contribution is exponentially suppressed
      in the large $N$-limit.  Later in the paper we provide further details on the structure of the partition function and the large $N$-limit.
      Thus there is hope that the present matrix model will provide the famous  $N^3$ dependence \cite{Klebanov:1996un}
       of the free energy in the large $N$-limit. For the matrix models arising from localization of 3D gauge theories \cite{Kapustin:2009kz,Jafferis:2010un,Hama:2010av,Hama:2011ea,Alday:2012au} the famous $N^{3/2}$ dependence of the free energy has been successfully demonstrated in different models, see for example \cite{Drukker:2010nc,Santamaria:2010dm,Herzog:2010hf,Jafferis:2011zi,Martelli:2011qj,Imamura:2011wg,Alday:2012au}. For a nice review, see \cite{Marino:2011nm}.

The paper is organized as follows: in section \ref{physics-S5} we review the construction from \cite{HosomichiSeongTerashima} and set the notations. Subsections \ref{vector}  and \ref{hyper-intro}  are devoted to the vector and hypermultiplets, respectively.
   In subsection \ref{local-gen} we review the localization argument and discuss the structure of the full
    partition function.  In section \ref{full-pert-calc} we discuss the actual calculation of the full perturbative partition function
     both for vector and hypermultiplets.
 We explain how to calculate the one-loop determinant using a change variables and  recasting the supersymmetry transformations into cohomological form. We then present the answer in terms of a matrix model.
      In section \ref{summary} a summary and open questions are presented. Moreover we give a possible interpretation of
       our matrix model.  Many technical details are presented in Appendices.

  Let us comment on the conventions which are used in this paper. As a main example we consider the unitary groups
   as gauge groups. For the Lie algebra  we follow mainly the conventions from \cite{HosomichiSeongTerashima},
    where the Lie algebra basis is defined in terms of Hermitian matrices  ($T_a^\dagger = T_a$) and the Killing form
     is positive definite ($\Tr (T_a T_b) = \frac{1}{2} \delta_{ab}$). For the various covariant derivatives (spinor, gauge, Levi-Civita etc) appearing in our formulae, we use the same symbol $D$, except in section \ref{hyper} and Appendix \ref{app-Killing}, where $\nabla$ is specifically reserved for the Levi-Civita connection.

\section{Supersymmetric theory on $S^5$}
\label{physics-S5}

The minimal 5D spinor representation for  Minkowski signature is four-dimensional and pseudoreal. The minimal supersymmetry algebra
 is generated by two charges which is a doublet with respect to the $SU(2)_R$ symmetry. This $SU(2)_R$ is an automorphism
  of the supersymmetry algebra.   The massless representations of this minimal supersymmetry  algebra are the vector
   multiplet and the hypermultiplet. The situation is similar for Euclidean signature and the spinor representation
    is still four-dimensional and pseudoreal.  In Appendix \ref{A-spinors} we collect our conventions for Euclidean 5D spinors.

In  this section we review the results from  \cite{HosomichiSeongTerashima}.  We briefly discuss the construction on $S^5$
of the minimal supersymmetric  5D theory  for vector and hypermultiplets. The theory on $S^5$ can be thought of as one-parameter deformations  of the flat Euclidean 5D model.

\subsection{Vector multiplet}
\label{vector}

The 5D vector multiplet contains a gauge field $A_m$, a real scalar $\sigma$ and a $SU(2)_R$-doublet of gauginos $\lambda^I$.
 We also need to introduce the auxiliary real fields $D_{IJ}$ with $D_{[IJ]}=0$ which form a triplet of $SU(2)_R$.  The spinor $\lambda^I$
  is in a real representation of $Spin(5) \times SU(2)$ (i.e.\ a $SU(2)$-Majorana spinor).
   It is well known how to write the $N=1$ supersymmetric Yang-Mills theory on $\mathbb{R}^5$.  However here we are interested in a deformation
   of this theory, namely in the supersymmetric theory on the five-dimensional sphere $S^5$ with radius $r$.
   The supersymmetry transformations are defined by (see \ref{spinor_bi_linear} for our notation of spinor bilinears)
\bea
&&\gd A_m = i\xi_I\Gc_m\gl^I~,\nn\\
&&\gd\gs = i\xi_I\gl^I~,\nn\\
&& \gd\gl_I = -\frac12(\Gc^{mn}\xi_I)F_{mn}+(\Gc^m\xi_I)D_m\gs-\xi^JD_{JI}+\frac{2}{r} t_I^{~J}\xi_J\gs~, \label{susy_vect} \\
&&\gd D_{IJ} = -i\xi_I\Gc^mD_m\gl_J+[\gs,\xi_I\gl_J]+\frac{i}{r}t_I^{~K}\xi_K\gl_J+(I\leftrightarrow J)~,\nn
\eea
where $\xi_I$ is a  spinor, satisfying the Killing equation on $S^5$
\bea
 D_m\xi_I=\frac{1}{r} t_I^{~J}\Gc_m\xi_J~,~~~t_I^{~J}=\frac{i}{2}(\sigma_3)_I^{~J}~,\label{killing_eqn}
 \eea
 where $\sigma_3=\rm{diag}[1,-1]$. In principle $t_I^{~J}$ can be chosen as any one of the three Pauli matrices.
The Lagrangian density on $S^5$ is defined as follows:
\bea&& L_{vector}= \frac{1}{g_{YM}^2} \Tr\Big[\frac{1}{2}F_{mn} F^{mn} -D_m\gs  D^m\gs-\frac12 D_{IJ}D^{IJ}+\frac{2}{r} \gs t^{IJ}D_{IJ}- \frac{10}{{r}^2}
 t^{IJ}t_{IJ}\gs^2\nn\\
&&\hspace{2cm}+i\gl_I\Gc^mD_m\gl^I-\gl_I[\gs,\gl^I]-\frac{i}{r}t^{IJ}\gl_I\gl_J\Big]~,\label{action_vector}
\eea
 where $F_{mn}$ is the field strength for $A_m$ and we use the standard $S^5$-metric for raising the indices.
   The $SU(2)_R$-indices are raised using
  $\epsilon^{IJ}$ (see Appendix  \ref{A-spinors}).
 The claim in \cite{HosomichiSeongTerashima} is that the corresponding action on $S^5$ is invariant
 under the transformation (\ref{susy_vect}) provided that the conditions
  (\ref{killing_eqn}) are satisfied. Here we use the self-evident notation for the covariant derivative $D_m$ which includes the gauge field
   and spin connection depending on which objects it acts. Under a gauge transformation $A_m$ transforms as a connection and all
    other fields are in the adjoint.  Here we regard $\xi^I$ as an even spinor and thus the supersymmetry transformations\footnote{It is a matter of convention which parameters to use in the supersymmetry transformations, even or odd spinors. The canonical convention is to use Grassmann odd spinor
     parameters for rigid supersymmetric theories and thus making the transformations even.} $\delta$ in (\ref{susy_vect})
     are odd.

The present theory on $S^5$ can be thought of as a deformation of the flat theory. In all formulas the radius $r$ can be sent consistently to
 infinity and we recover the corresponding formulas on $\mathbb{R}^5$.

\subsection{Hypermultiplet}
\label{hyper-intro}

Next let us discuss the $N=1$ matter multiplet (hypermultiplet)  in $5D$ as formulated in \cite{HosomichiSeongTerashima}.
The field content of the hypermultiplet consists of a pair of complex scalars $q^A_I,~~I=1,2$ and a fermion $\psi^A$, with the reality conditions
\bea
 (q^A_I)^*=\Omega_{AB}\epsilon^{IJ}q^B_J~,~~(\psi^A)^*= \Omega_{AB}C\psi^B~,\label{reality-cond-spin}
 \eea
where $\Omega_{AB}$ is the invariant tensor of $Sp(N)$ and $C$ is the charge conjugation matrix and thus the index $A$ runs from $1$ till $2N$.
  The field $q^A$ is a doublet of $SU(2)_R$ and $\psi^A$ is a singlet of $SU(2)_R$.

One can minimally couple the hypermultiplet to the vector multiplet by gauging a subgroup of $Sp(N)$. As an example let us consider the case of  $SU(N)$  gauge group. We embed $SU(N)$ into $Sp(N)$ in the standard manner (by viewing $Sp(N)$ as $N\times N$ anti-Hermitian quaternion matrices)
\bea U\to \left|\begin{array}{cc}
            U & 0 \\
            0 & U^{-T}
          \end{array}\right|~,~~~U\in SU(N),~~~\Go=\left|\begin{array}{cc}
                                                          0 & 1 \\
                                                          -1 & 0
                                                        \end{array}\right|\nn
                                                        \eea
and one can rewrite the scalar field $q$ into a more familiar form as
\bea q_1=\frac{1}{\sqrt2}\left|\begin{array}{c}
           \phi_+ \\
           \phi_-
         \end{array}\right|~,~~~q_2=\frac{1}{\sqrt2}\left|\begin{array}{c}
           -\phi_-^* \\
           \phi_+^*
         \end{array}\right|~,\label{rewrite_scalar}\eea
where $\phi_{\pm}$ transform in the $\bf{N}$ and $\bar{\bf{N}}$ of $SU(N)$, respectively. The fermion can be written in a similar
 manner
\bea \psi^A=\frac{1}2\left|\begin{array}{c}
                    \psi^\alpha \\
                    -C\psi^*_\beta
                  \end{array}\right|~,\label{rewrite_fermion}\eea
where $\psi^\alpha$ is now an unconstrained Dirac spinor transforming in $\bf N$ (here $\alpha$ is the index for the representation).
  Analogously we can discuss the adjoint representation of
 $SU(N)$ when two copies of the adjoint are embedded into that of $Sp(N)$.

Suppressing the $A$-index the supersymmetry transformations are defined as follows:
 \bea
 &&\delta q_I=-2i\xi_I\psi~,\nn\\
&&\delta\psi=\Gc^m\xi_I(D_mq^I)+i\sigma \xi_Iq^I-\frac{3}{r} t^{IJ}\xi_Iq_J~, \label{hyper-tran-noaux}
\eea
and provided that $\xi_I$ satisfies the Killing spinor
 equation (\ref{killing_eqn}) these transformations leave invariant the action with the following
  Lagrangian density
\bea
&&L_{matter}=\epsilon^{IJ}\Omega_{AB}D_mq_I^A  D^m q_J^{B} -\epsilon^{IJ}q_I^A \gs_{AC} \gs^C_{~B}  q_J^B +
\frac{15}{2r^2} \epsilon^{IJ}\Omega_{AB}t^2  q_I^A  q_J^B
 \nn\\
&&\hspace{1cm}-2i \Omega_{AB}\psi^A\slashed{D}\psi^B-2\psi^A\gs_{AB} \psi^B -4\Go_{AB}\psi^A\gl_Iq^{IB}-iq_I^AD_{AB}^{IJ}q_J^B~,\label{action-matter-1}
\eea
where $t^2=t^{IJ}t_{IJ}=1/2$ and $\sigma_{AB} = \Omega_{AC} \sigma^C_{~B}$.
Using  (\ref{rewrite_scalar}) and  (\ref{rewrite_fermion}) we can rewrite the action (\ref{action-matter-1})
 in a more conventional form
\bea
L_{matter} = (D_m \phi_+)^{\dagger}(D^m \phi_+)+(D_m \phi_-)^{\dagger}(D^m \phi_-)-i (\psi^\alpha)^{\dagger}\slashed{D}\psi^\alpha+\cdots ~, \label{action-matter-2}
\eea
 which is more familiar when we discuss the relation to the $N=2$ 4D hypermultiplet.

  Later on we will analyze the case when the hypermultiplet is in a representation $R$.  If we consider the special case when the hypermultiplet is in the adjoint representation, the theory defined by
 \bea
  L_{vector} + L_{matter}\label{N=2deformed}
 \eea
in the limit $r \rightarrow \infty$ becomes $N=2$ 5D super Yang-Mills. This is a maximally supersymmetric
  Yang-Mills theory which can be obtained from reducing  the ten dimensional supersymmetric Yang-Mills action.
   The five scalars are the field $\sigma$ coming
   from the vector multiplet plus four real scalars coming from the two complex scalars $\phi_\pm$ in the hypermultiplet. The scalar $\sigma$
    corresponds to the reduction along the time direction of the ten dimensional theory and this is why the kinetic term for $\sigma$ in (\ref{action_vector}) has
     the wrong sign as compared to the $F^2$-term.
       As before we think about (\ref{N=2deformed}) as a deformation of the $N=2$ 5D supersymmetric Yang-Mills, with the deformation controlled
      by the parameter $r$.

  In order to calculate the partition function using the localization technique, we need an off-shell version of the supersymmetry transformations (\ref{hyper-tran-noaux}).
   In subsection \ref{hyper} we present the off-shell version of the supersymmetry algebra written in new variables.
    At this point we deviate from the supersymmetry transformations
  presented in \cite{HosomichiSeongTerashima} since we think that there are issues of global nature related to their off-shell construction.

\subsection{Localization}
\label{local-gen}

We are interested in calculating the partition function on $S^5$ of the deformed $N=1$
 Yang-Mills theory coupled to matter hypermultiplets using localization techniques.  If
  we take the hypermultiplet in the adjoint representation we will refer to the theory as
  $N=2$ supersymmetric Yang-Mills theory (to be more precise a deformation of the flat theory
   preserving 8 supercharges).
   In \cite{HosomichiSeongTerashima} the localization locus for these theories has been
  discussed although no actual calculation has been presented and it is our goal to present
   a concrete calculation of the perturbative partition function for these models.

  The action for $N=1$ supersymmetric Yang-Mills is given by
  \bea
   S_{N=1} = \int\limits_{S^5} d^5 x \sqrt{g} ~L_{vector}
  \eea
  and the action for $N=1$ supersymmetric Yang-Mills with matter is given by
  \bea
   S_{N=2} = \int\limits_{S^5} d^5x \sqrt{g} ~(L_{vector} + L_{matter})~.
  \eea
   As we mentioned above, if we take the matter hypermultiplet in the adjoint then we call the theory $N=2$ supersymmetric Yang-Mills.
    The partition function for  $N=1$ supersymmetric Yang-Mills is
    defined by
   \bea
    Z = \int e^{-S_{N=1}}~,
   \eea
where the minus sign is chosen to have a Gaussian damping for the $F^2$-term in (\ref{action_vector})    (remember our conventions for the trace). The partition function for $N=1$ supersymmetric Yang-Mills with matter is defined
      with the same choice of sign. Following the standard arguments we have to add to the action the term $\delta V$
      \bea
        Z(s) = \int e^{-S_{N=1} + s ~\delta V}~,
      \eea
      such
    that from one side $\delta^2 V =0$ and from  the other side $\delta V$ is positive. Using these two conditions we can argue that
     $Z(s)$ is independent of $s$ and
     only the fixed points of $\delta$ together with the one-loop determinants contribute to the path integral.

 As discussed in \cite{HosomichiSeongTerashima}, for $S_{N=1}$ we can add the following
  term which fulfills the above requirements:
  \bea
 \delta V = \delta  \int\limits_{S^5} d^5x \sqrt{g}~ \Tr [ (\delta \lambda_I)^\dagger \lambda_I ]~. \label{exact-vector-term}
  \eea
  Provided that we integrate over imaginary $\sigma$ we get the following localization locus:
  \bea
   \iota_v * F =  F~,~~~~~~D \sigma=0~~~~~~D_{IJ}=-2t_{IJ}\sigma~,\label{local-locus-vect}
  \eea
   which is understood modulo gauge transformations. Also note that the first condition above implies $\iota_vF=0$.
   The choice of imaginary $\sigma$ also fixes the problem with the sign in front of the kinetic term for $\sigma$ in
    (\ref{action_vector}).  We remark that the localization locus $\eqref{local-locus-vect}$ is the same as for the 5D topological field theory constructed and studied in \cite{KallenZabzine12}.  On the solutions of (\ref{local-locus-vect}) the bosonic
      part of the action is given by
  \bea
   S_{N=1} = \frac{1}{g_{YM}^2} \int\limits_{S^5}  ~ \Tr ( F \wedge * F  +d^5 x \sqrt{g}\frac{8}{r^2} \sigma^2)~,\label{class-vect-act}
  \eea
where 
    we remember that we integrate over imaginary $\sigma$ (so we have sent $\sigma$ to $i\sigma$). Both terms
    in (\ref{class-vect-act}) are positive and the $F^2$-term is zero if and only if $F=0$. On $S^5$ the condition $F=0$ implies that $A$ is gauge equivalent to 0 and $\sigma$ is constant. Indeed on $S^5$ the solution $A=0$ and $\sigma = \rm{constant}$ is an isolated point in the space of gauge equivalence classes of solutions to  (\ref{local-locus-vect}).
    All other solutions of the equations (\ref{local-locus-vect}) will have non-zero $F^2$-term and it will be related
     to the instanton number on $\mathbb{C}P^2$ as we will explain in section \ref{moreoncontactinstantons}. Thus the partition function will have
      the following schematic form:
  \bea
   Z= ({\rm contribution}~{\rm of}~A=0 ~{\rm and}~ \sigma =\textrm{constant}) + \sum\limits_{n\neq 0} e^{-\alpha \frac{r}{g_{YM}^2} n} ( \cdots)_n~,
   \label{part-funct-schem}
  \eea
         where $n$ is the integral instanton number for $\mathbb{C}P^2$ and $\alpha$ is some positive numerical constant.
         The term $( \cdots)_n$ stands for the contribution of the one-loop determinant around instantons of fixed instanton number and in general it is
          quite hard to evaluate, especially for a compact space such as $\mathbb{C}P^2$.  If we consider the analog of the 't Hooft
           large $N$-limit
           \bea
            \lambda = \frac{g_{YM}^2}{r} N= {\rm fixed}~,~~~~~~N\rightarrow \infty~,
           \eea
                      then the terms with $n\neq 0$ are believed to be exponentially suppressed. Thus provided that the one-loop term denoted by $(\cdots)_n$ does not overcome the exponential suppression, we expect that
            in the large $N$ limit only the contribution from $A=0$ and $\sigma = {\rm constant}$ is essential.

  Next, if we want to discuss the localization for $N=1$ supersymmetric Yang-Mills with matter  then in addition to the term
    (\ref{exact-vector-term}) we have to add the following term for the hypermultiplet:
  \bea
  \delta \int d^5x \sqrt{g}~ \Tr [(\delta \psi)^\dagger \psi]~. \label{hyper-exact-term}
  \eea
   Assuming that $A=0$ and $\sigma={\rm constant}$ the corresponding localization locus for the bosonic fields is very simple
   \bea
    q_I=0~,~~~~~{\cal F}=0~,
   \eea
 as will be shown in detail in section \ref{hyper}.   Thus the partition function for the $N=1$ theory with matter and the $N=2$ theory has
    a similar form to (\ref{part-funct-schem}), but with additional contributions to the one-loop determinants from the hypermultiplet.

\section{Calculation of the one-loop determinants and more discussion about the localization locus}
\label{full-pert-calc}

In this section we will both discuss the localization locus further and calculate the full perturbative contribution coming from evaluating the one-loop determinants around
 $A=0$ and $\sigma ={\rm constant}$ for the $N=1$ model (and in addition the one-loop determinants around $q_I=0$, ${\cal F}=0$ for the model including the hypermultiplet). For the calculation of the one-loop determinants, in principle
  one can try to do it right away by the expanding up to quadratic order the expressions (\ref{exact-vector-term})
  and (\ref{hyper-exact-term}) and then calculating the appropriate determinants of the Laplace and Dirac  operators on $S^5$.
   This is how the calculation has been done for the theory on $S^3$ by Kapustin et al.\ \cite{Kapustin:2009kz}. However $S^3$ is a group
    manifold and the spectrum of the Laplace and Dirac operators can easily be worked out. On $S^5$ it would require more
     work. Alternatively we can follow a different path suggested by  Pestun \cite{Pestun:2007rz}. Namely we can
      make a change of variables in field space and recast the supersymmetry transformations into a cohomological form. Then the calculation
       requires the use of an appropriate index theorem. The cohomological version of the calculation for supersymmetric theories on $S^3$ (as well as on general Seifert manifolds) has been  performed in \cite{Kallen11} (see also \cite{Ohta:2012ev} for a treatment of matter fields in this approach and \cite{Benini:2012ui} for a similar calculation for 2D theories on $S^2$).  On $S^3$ obviously both approaches produce the same final result.
       For the calculation on $S^5$, we follow the approach suggested by Pestun and we rewrite the supersymmetry transformation in cohomological form
        which will allow us to calculate the one-loop determinants in an elegant way using the Atiyah-Singer index theorem for the twisted Dolbeault operator. Below we will focus on showing how to rewrite the supersymmetry transformations into a cohomological form. After this is done, much of the remaining calculations have already been performed in \cite{KallenZabzine12}, where from we will borrow many results.

\subsection{Constructing geometrical quantities using Killing spinors}

As a preparation for the actual calculation let us first discuss some geometry entailed by the existence of two normalized Killing spinors. The parameter $\xi_I$  in the supersymmetry transformations  is  a \emph{bosonic} (even) Killing spinor satisfying equation (\ref{killing_eqn}) and normalized as follows
\bea
\xi_I\xi_J=-\frac{1}{2} \epsilon_{IJ}~.\label{normalize}
\eea
For $S^5$, the explicit solution for $\xi_I$ was given in \cite{HosomichiSeongTerashima}, and we  also present the solution in terms of differential forms in Appendix \ref{A-KIll-spin}. The fact that the five-manifold admits two Killing spinors has some simple but profound implications on the property of the manifold. Even though most of these properties are well known, we include a list of those in Appendix \ref{app-Killing} as well as a sketch of their derivation. Using the Killing spinors we can define a nowhere vanishing
 vector field $v^m=\xi_I\Gc^m\xi^I$. In the case of $S^5$, with the choice of $\xi_I$ in \cite{HosomichiSeongTerashima} the vector field generates the $U(1)$-action in the Hopf fibration
\be
\begin{array}{lll}
  S^{5} & \longleftarrow & S^1\\
{\scriptstyle \pi}\Big\downarrow && \\
  {\mathbb C}P^{2} &   &
\end{array}\label{bundle-general-BW}
\ee
 The corresponding 1-form $\kappa_m = g_{mn} v^n$ is a connection for this Hopf  fibration and it
  defines a contact structure, $\kappa \wedge (d\kappa)^2 \neq 0$. Moreover $-d\kappa$ coincides with
 the pull-back of the standard symplectic form
  associated to the Fubini-Study K\"ahler metric on ${\mathbb C}P^{2}$ (this can be derived from the relations (\ref{dkappa}) and    (\ref{hor_complex_structure})).
   For further details regarding the contact geometry involved in this case, the reader may consult \cite{KallenZabzine12}
   and the references  therein.

  In general, the existence of two normalized Killing spinors on a five-manifold implies that we deal with a $K$-contact structure.
   A wide class of examples is given by $U(1)$-fibrations over   four-dimensional
   K\"ahler manifolds with an integral K\"ahler  form (a so called Hodge manifold).
    More conditions should be imposed on the Hodge manifold in order
     to guarantee the existence of globally defined Killing spinors on the five-dimensional manifold.
      For a recent discussion on related issues in three and four dimensions
      one may consult \cite{Festuccia:2011ws,  Klare:2012gn,  Dumitrescu:2012ha}.  For the 5D case a similar analysis needs to be performed in order to establish on which backgrounds 8 supercharges exists.

Let us finish the geometrical discussion with some terminology (see \cite{KallenZabzine12} for more details).
 Since we are dealing with a $S^1$ fibration with a choice of connection 1-form $\kappa$ we can introduce a decomposition
  of 2-forms $\Omega^2$ into the vertical part $\Omega^2_V$ and the horizontal part $\Omega_H^2$. In turn, the horizontal
   part $\Omega^{2}_H$ can be decomposed into horizontal self-dual and anti-self-dual parts defined by the projectors
   $\frac{1}{2}(1 \pm \iota_v *)$. Thus we have the decomposition of 2-forms into the spaces
\bea
 \Omega^2 (S^5) = \Omega^2_V (S^5) \oplus \Omega^{2+}_H (S^5) \oplus \Omega^{2-}_H(S^5)~,
\eea
 which are orthogonal to each other with respect to the standard scalar product defined by the metric.  Here we would like to
  point out that in the present context what we call horizontal self-dual or horizontal anti-self-dual is relative and it depends on the conventions chosen.
  By changing $v\rightarrow -v$
   and $\kappa \rightarrow - \kappa$ and keeping the metric unchanged we exchange the notion horizontal self-duality and horizontal
    anti-self-duality. Our choice corresponds to $\kappa \wedge (d\kappa)^2$ being minus the volume form given
     by metric.

\subsection{Vector multiplet}
\label{sec_cov}

 In this subsection we will discuss the localization locus further and calculate the contribution from the vector multiplet, or in other words the full perturbative
 partition function of  $N=1$ supersymmetric Yang-Mills. Following Pestun's ideas \cite{Pestun:2007rz}
  it is convenient to bring the supersymmetry  transformations (\ref{susy_vect}) to cohomological form
   by a change of variables for fermonic fields. This change does not generate any field dependent Jacobian in the path integral.

\subsubsection{Cohomological form of supersymmetry transformations}
Using the Killing spinors $\xi_I$, we can convert $\gl_I$ to a 1-form and a 2-form:
\bea \Psi_m=\xi_I\Gc_m\gl^I~,~~~~~\chi_{mn}=\xi_I\Gc_{mn}\gl^I-\kappa_{[m}   \xi_I\Gc_{n]}\gl^I~.\label{field_redef}
\eea
While $\Psi$ is an unrestricted 1-form, the 2-form $\chi$ satisfies the following conditions:
\bea \iota_v\chi=0~,~~~~\iota_v * \chi=-\chi~,
\eea
meaning that $\chi$ is a horizontal anti-self-dual 2-form.

The formula (\ref{field_redef}) can be inverted to write $\gl_I$ as
\bea\gl_I=-\frac12\xi^J\Gt^{mn}_{JI}\chi_{mn}+(\Gc^m\xi_I)\Psi_m~,\label{redefine_spinor_fine}
\eea
where $\Theta^{IJ}_{mn}=\xi^I\Gc_{mn}\xi^J$ (see Appendix \ref{app-Killing} for more properties).
The fermion $\gl_I$ has 8 real components which is the same as the
5 components of  $\Psi$  plus  3 more from $\chi$.

In these new odd variables the supersymmetry transformations (\ref{susy_vect}) can be rewritten as follows
\bea
&&\delta A_m = i\Psi_m~,\nn\\
&& \gd \gs = i v^m \Psi_m~,\nn\\
&& \gd \Psi_m = v^n F_{nm}+D_m \gs~,\label{susy_vect_twist}    \\
&& \gd \chi_{mn} = H_{mn}~,\nn\\
&&\gd H_{mn}= i\L^A_v\chi_{mn}-[\gs,\chi_{mn}]~,\nn
\eea
 where $\L^A_v = \L_v + i [~,\iota_v A] $.
 Here the 2-form $H$ is defined as
\bea
H =(1-\iota_v * )F-(\kappa\wedge\iota_vF)+\Gt^{IJ}(D_{IJ}+2t_{IJ}\gs)~,\nn
\eea
 where $\Theta^{IJ}$ is the 2-form defined in (\ref{dkappa}).  Like $\chi$, $H$ is a horizontal anti-self-dual 2-form.
 The square of the transformations (\ref{susy_vect_twist}) is given by
\bea
  \delta^2 = i \L_v + G_{i(\sigma -\iota_v A )} ~,
 \eea
 where $\L_v$ is the Lie derivative along $v$ and $G_{i(\sigma -\iota_v A )}$ is a gauge transformation with parameter
  $i (\sigma -\iota_v A )$. With our conventions the gauge field is transformed as
\bea
 G_{\epsilon} A = d\epsilon - i [A, \epsilon]
\eea
and on all other fields in the adjoint as
\bea
G_{\epsilon} \bullet  = -i [\bullet    ,\epsilon ]~.
\eea
As explained before we have to send $\sigma$ to $i\sigma$ and in the
new variables the $\gd$-exact term can be written as follows:
\bea \delta V = \delta \int \text{Tr} \big (\chi \wedge * (2F_H^- - H) + \frac12\Psi \wedge * \gd\overline{\Psi} \big )~,
\eea
 where $F_H^-$ is the horizontal anti-self-dual part of the field strength ($\iota_v * F^-_H = - F^-_H$). The bosonic part
  of the above term is zero if
  \bea
   F_H^-=0~,~~~~~~\iota_v F =0~,~~~~~~~D\sigma=0~.\label{general-vect-loc1}
  \eea
  These equations describe the localization locus for the path integral.

  \subsubsection{Localization locus on contact instantons} \label{moreoncontactinstantons}

  The conditions (\ref{general-vect-loc1}) can be combined as in (\ref{local-locus-vect}), which can be alternatively written as
  \bea
   * F = \kappa \wedge F~,~~~~~ D_A\sigma=0~, \label{loc-vect-123}
  \eea
 where we use a subscript on $D_A$ to indicate with which connection we form the covariant derivative.
   The first equation has been introduced in the context of topological 5D Yang-Mills theory in \cite{KallenZabzine12}. This equation can be
    written on any contact five-manifold and we refer to this equation  as a contact instanton (these equations have been discussed in the recent work \cite{Wolf:2012gz}, see also \cite{Harland:2011zs}  for a related system of equations, and also more references therein).
      Here our goal is
     to study this equation specifically on $S^5$.

   One can show that the first equation in (\ref{loc-vect-123}) implies the Yang-Mills equation
   \be
    D_A(* F ) = d\kappa \wedge F =0~,
   \ee
   since $\iota_v * d\kappa = - d\kappa$, which is proved using
     the relations from Appendix \ref{app-Killing}.   Indeed this is a non-trivial fact, for example the equation with different sign, $* F =- \kappa \wedge F$,  does not   imply the Yang-Mills equation.  For gauge group $U(1)$, the contact instanton equation implies that $F\in H^2(M, \mathbb{R})$
      and thus on $S^5$ the flat connection $F=0$ ($A=0$ up to gauge transformations) is the only contact instanton.

To explore  further the contact instanton equation for the non-abelian case, we need to go to a convenient gauge. From $\iota_vF =0$ we derive
\bea
0=\iota_vF=\iota_vdA-i[\iota_vA,A]=\L_vA-D_A(\iota_vA)~.\nn\eea
 This shows that the Lie derivative of $A$ along $v$ is a gauge transformation and thus we can choose the gauge
\bea \L_vA=0~.\label{gauge_choice}\eea
We will define the vertical and horizontal part of the connection $A$ as
\bea
 \varrho=\iota_vA~,~~~\ga=A_H=A-\kappa\iota_vA \nn\eea
to save on the use of subscripts. Since both $\varrho$ and $\ga$ are basic with respect to
 the $U(1)$-action generated by the vector field $v$ (meaning horizontal and invariant), they can be pushed down to the base $\BB{C}P^2$. Furthermore, under a gauge transformation $\delta_{\rm gauge} A=D_Af$ with $f$ independent of the circle direction
   (which preserves the gauge (\ref{gauge_choice})),
\bea
\delta_{\rm gauge} \varrho=-i[\varrho,f]~,~~~\gd_{\rm gauge} \ga=D_{\ga}f~,\nn\eea
i.e. $\ga$ and $\varrho$ can be regarded as a gauge connection and an adjoint scalar on $\mathbb{C}P^2$.
Due to $\L_v\ga=0$ we have that $d\ga$ is basic automatically.

Expanding out $F$ in $\varrho$ and $\ga$
\bea F=d\ga+(d\kappa)\varrho-\kappa\wedge  d\varrho-i\ga\wedge \ga-i\kappa\wedge [\varrho,\ga],\label{F_expand}\eea
we further get $0=\iota_vF=-d\varrho-i[\varrho,\ga]=-D_{\ga}\varrho$, i.e.\ $\varrho$ is a covariantly constant scalar on $\mathbb{C}P^2$.

Now we explore the implication of the first equation in (\ref{loc-vect-123}). Again using the explicit form of (\ref{F_expand}), we get
\bea &&
F_H^-=\big(d\ga+(d\kappa)\varrho-i\ga\wedge \ga\big)^-=(d\kappa)\varrho+F(\ga)^-~,\nn\\
&&F_H^+=\big(d\ga-i\ga\wedge\ga\big)^+=F(\ga)^+~,\label{temp_2}\eea
where $F(\ga)$ is the curvature on $\BB{C}P^2$ of the connection $\ga$. We point out once again that, due to our unfortunate choice of $\kappa$, we have that $\kappa \wedge d\kappa \wedge d\kappa$ determines an orientation opposite to that of $\sqrt{g}dx^1\wedge \cdots\wedge dx^5$, and hence what gets a $+$ superscript is actually anti-self dual from the 4D point of view while those with $-$ are self-dual. In particular, $d\kappa$ is minus the K\"ahler form of $\BB{C}P^2$ and hence self-dual.

Finally, we have reduced our contact instanton condition into a pair of equations on $\BB{C}P^2$
\bea D_{\ga}\varrho=0~,~~~F(\ga)^-=-\varrho d\kappa~,\label{Donaldson}
\eea
which is a perturbed version of the Donaldson equation $F(\ga)^-=0$. Still one must bear in mind the 5D origin of these equations and there is still the possibility of a large $\theta$-dependent, where $\theta$ is the coordinate along the $S^1$ fibre, gauge transformation that preserves (\ref{gauge_choice}).

Since by construction $D_{\ga}F(\ga)=0$, together with $D_{\ga}\varrho=0$ we have
\bea
D_{\ga}F(\ga)^-=0~,\nn\eea
which implies that $F(\ga)$  satisfies the Yang-Mills equation on $\BB{C}P^2$: $D_{\ga}^{\dagger}F(\ga)=0$.
The second  equation in (\ref{Donaldson}) is a perturbation of the Donaldson equation in the following sense. First, if $\ga$ is an \emph{irreducible} connection, then $\varrho=0$, and the perturbation vanishes. Assume that $\ga$ is a reducible connection, which means that the holonomy group of $\ga$ defines a conjugacy class of proper subgroups of $SU(N)$. Take any point $m\in\BB{C}P^2$, assume that the holonomy group at this point is $H\subset SU(N)$, then by definition any $h\in H$ is realized as the holonomy of a loop based at $m$. The fact that $\varrho$ is a parallel scalar implies
\bea
h^{-1}\varrho h=\varrho~,~~\forall h~,\nn\eea
by using the second equation in (\ref{Donaldson}), one gets $h^{-1}F(\ga)^{-} h=F(\ga)^{-}$, $\forall h\in H$. First assume that $H$ is a semi-simple subgroup of $SU(N)$, then one can conclude $F(\ga)^{-}=0$ from the semi-simplicity.

What remains is the case when $H$ is $U(1)$. We deduce from $D_{\ga}\varrho=0$ that $\Tr[\varrho^2]$ is a constant, which we assume to be non-zero.
Furthermore, from $D_{\ga}F(\ga)^+=0$ we have $D_{\ga}\Tr[\varrho F(\ga)^+]=0$ and
\bea
 D^{\dagger}_{\ga}\Tr[\varrho F(\ga)^+]=\Tr[*(D_{\ga}\varrho)\wedge*F(\ga)^+]+\Tr[\varrho D^{\dagger}_{\ga}F(\ga)^+]=0~.\nn
\eea
This shows that $\Tr[\varrho F(\ga)^+]$ is a harmonic 2-form with negative self-intersection number, which does not exist for $\BB{C}P^2$.
We have thus $\Tr[\varrho F(\ga)^+]=0$, then as $\varrho\neq0$ by assumption and that $\varrho$ and $F(\ga)$ are embedded into the same generator of $SU(N)$,
we conclude $F(\ga)^+=0$.
Thus we see that $F=0$ on $S^5$ and hence $A$ is gauge equivalent to zero.

To summarize the foregoing strenuous argument, we conclude that the perturbation in (\ref{Donaldson})
 can be ignored in all cases except for the reducible $U(1)$ instantons, which are gauge
 equivalent to trivial connections. One gets finally the value of the Yang-Mills kinetic term on contact instantons
\bea
&&\int\limits_{S^5}~\Tr[F\wedge* F]\Big|_{\textrm{cnct-inst}}=\int\limits_{S^5}\kappa\wedge \Tr[F_H^+\wedge F_H^+]
\stackrel{\textrm{push down}}{=}2\pi r\int\limits_{\BB{C}P^2}\Tr[F_H^+\wedge F_H^+]\nn\\
&=&2\pi r\int\limits_{\BB{C}P^2}\Tr[F(\ga)^+\wedge F(\ga)^+]
=2\pi r\int\limits_{\BB{C}P^2}\Tr[F(\ga)\wedge F(\ga)]=16 \pi^3 r \BB{Z}_+~.\eea
 Exactly this fact we have used in our discussion of the full partition function  (\ref{part-funct-schem}).

\subsubsection{Matrix model for the vector multiplet}

 Now we want to calculate the one-loop determinant around the zero connection and constant $\sigma$.
Modulo some conventions the calculation is totally identical to the one performed in \cite{KallenZabzine12}. Therefore we will here only explain the philosophy behind the calculation and sketch the derivation with the present conventions taken into account. We refer the readers to
  \cite{KallenZabzine12} for further explanations and technical details. The philosophy behind the calculation is the following. In analogy with equivariant localization of finite-dimensional integrals, \cite{Atiyah:1984px,MR705039,Schwarz:1995dg}, we can argue \cite{Pestun:2007rz} that the one-loop determinant will be given by evaluating the superdeterminant of
   the operator
\be
 i\mathcal{L}_v -i [\sigma,~]\label{oper-vect-1loop}
\ee
acting on the spaces which are given by the 'coordinates' in the transformations $\eqref{susy_vect_twist}$. In these transformations, the supersymmetry transformation can be understood as an equivariant differential acting on a superspace with coordinates given by the gauge field $A$ and the odd anti-self-dual field $\chi$. After including ghost fields needed for gauge fixing in the transformations $\eqref{susy_vect_twist}$, it is shown in \cite{KallenZabzine12} that the one-loop determinant is given by the ratio of determinants of the operator $\eqref{oper-vect-1loop}$ acting on even and odd, respectively, horizontal anti-holomorphic differential forms.
Here we assume that the range of the coordinate $\theta$ in the fibre direction is $0\leq \theta<2 \pi r$, which means that the eigenvalues of $\mathcal{L}_v$ is $-\frac{i t}{r}, ~t\in\mathbb{Z}$ (with chosen conventions the vector $v$ corresponds to $-\partial_\theta$). The imaginary $i$ in (\ref{oper-vect-1loop}) comes from the integration over imaginary $\sigma$ and we
 use the conventions with Hermitian Lie algebra generators. Taking this into account the one-loop contribution is written
as follows
\bea
 Z_{\rm 1-loop}^{\rm vect} =\prod\limits_\beta\prod\limits_{t \neq 0}\left( \frac{ t}{r} - \langle \beta,i\sigma\rangle   \right)^{(1+\frac{3}{2}t+\frac{1}{2}t^2)}~,\label{vect1-loop-beg}
\eea
 where $\beta$ stands for the roots of the Lie algebra. This expression arises as follows. For each element in the space $\Omega_{H}^{0,\bullet}$, the operator $\eqref{oper-vect-1loop}$ acts with the same eigenvalue on both the even and the odd forms. The modes labelled by $t$ is a section of the bundle ${\cal O}(t)$ over $\BB{C}P^2$, and the cancelation between the numerator and denominator is determined by the index of the Dolbeault complex twisted by this line bundle. This is how the exponent in $\eqref{vect1-loop-beg}$ arises, it is the number of left over modes after the cancelation has taken place, for each $t$. Again, for further explanations and technical details, we refer to \cite{KallenZabzine12}.

    Rewriting (\ref{vect1-loop-beg}) as a product over positive roots only
 we get
\be
 Z_{\rm 1-loop}^{\rm vect}  = \prod\limits_{\beta > 0}
  \prod\limits_{t\neq 0}\left( \frac{ t^2}{r^2} -  (\langle \beta,i\sigma\rangle)^2   \right)^{(1+\frac{3}{2}t+\frac{1}{2}t^2)}
  =\prod\limits_{\beta > 0}
  \prod\limits_{t=1}^\infty \left( \frac{ t^2}{r^2} -  (\langle \beta,i\sigma\rangle)^2   \right)^{(2+t^2)}~.
\ee
 Using the infinite products (\ref{infin-exp-sin}) and (\ref{our-func-exp}) the one-loop contribution can be written as
\be
 Z_{\rm 1-loop}^{\rm vect}  = \prod\limits_{\beta > 0} (2\pi r e^{-\zeta'(-2)} )^2  \left ( \frac{\sin (\pi \langle \beta, i r\sigma \rangle)}
  {\pi \langle \beta, i r\sigma \rangle} \right )^2 e^{f(\langle \beta, i r\sigma \rangle)}~,\label{final-1lop-vect}
\ee
 where the function $f$ is defined in (\ref{definition-f-func}).
The Yang-Mills term $\eqref{class-vect-act}$ evaluated on the localization locus $A=0$ and $\sigma ={\rm constant}$ gives us
\begin{equation}
S_{N=1}= \frac{8}{g_{YM}^2}\text{vol}(S^5)\text{Tr}\left(\frac{\sigma}{r}\right)^2= \frac{ 8 \pi^3 r}{g_{YM}^2}\text{Tr}\left( r\sigma\right)^2~.
\end{equation}

 Summarizing everything and
introducing the dimensionless combination $\phi =  r \sigma$ the final answer can be written
as a matrix integral over the Cartan subalgebra
\be
&& Z
  =\int\limits_{\rm Cartan} [d\phi]~e^{- \left [ \frac{ 8\pi^3 r}{g_{YM}^2}  \right ] \text{Tr}(\phi^2)} \prod_{\beta >0}
    \sin^2 ( \pi \langle\beta , i\phi\rangle ) \cdot e^{ f( \langle \beta, i \phi\rangle )}  \nn \\
&& =\int\limits_{\rm Cartan} [d\phi]~e^{- \left [ \frac{ 8\pi^3 r}{g_{YM}^2}  \right ] \text{Tr}(\phi^2)}  {\rm det}_{\rm Ad}
  \left (    \sin ( i\pi \phi ) \cdot e^{\frac{1}{2} f(i \phi )} \right ) ~,  \label{vector1loop}
\ee
where we ignore all irrelevant numerical factors.   The factor $(\langle \beta, i r\sigma \rangle )^{-2}$ in (\ref{final-1lop-vect})
 is canceled
 due to pushing the integration from the whole Lie algebra to the Cartan subalgebra.
 Let us comment on properties of the matrix model (\ref{vector1loop}). The potential $f(ix)$ for $x$ real is a symmetric real (the branch cuts cancel between polylogs and logs) function, with the asymptotic behaviour
 \bea \lim_{x\to\pm\infty} f(ix)\sim -\frac{\pi}{3}|x|^3~,\nn\eea
and, as can be checked using Mathematica, the function $f$ behaves very nicely along
  the imaginary axis. Due to these facts, we believe that the matrix integral is a well-defined converging integral.

 In  addition to the $N=1$ Yang-Mills action we can add supersymmetric versions of the Chern-Simons couplings
     \bea
  i  \int\limits_{S^5} \text{Tr} ( \kappa \wedge F \wedge F) = \text{Tr} \int\limits_{S^5}
     \text{Tr} ( d\kappa \wedge (A \wedge d A + \frac{2}{3} A\wedge A \wedge A))
     \eea
     and
     \be
i  \int\limits_{S^5} {\rm Tr } \left (A\wedge dA \wedge dA + \frac{3}{2} A\wedge A \wedge A \wedge dA + \frac{3}{5} A\wedge
  A \wedge A \wedge A \wedge A \right )~,
\ee
 which would generate the terms $i \text{Tr} (\phi^2)$ and $i\text{Tr} (\phi^3)$ in the exponent, respectively.
  The rest of calculation will be unchanged when adding these terms.

\subsection{Hypermultiplet}
\label{hyper}

Next we will evaluate the contribution of the hypermultiplet to the perturbative partition function.  We assume that the
 hypermultiplet is in a representation $R$ of the gauge group.

 Using the Killing spinors  $\xi_I$  we can combine $q_I$ with $\xi_I$  and define a new bosonic spinor field $q$
\bea
 q=\xi_Iq^I~,~~~~~q_I=-2\xi_I q~.\nn
 \eea
From the reality condition satisfied by $\xi_I$ and $q_I$ one can see that the spinor field $q$ now satisfies the same reality condition as $\psi$.

We can rewrite the supersymmetry transformation (\ref{hyper-tran-noaux}) in terms of $q$ and $\psi$ and
these transformations close only on-shell.  However for the sake of localization, we need an odd symmetry that squares off-shell to a translation plus a gauge transformation for \emph{one} set of specifically chosen parameters. To obtain this, we first use the vector field $v$ to define a chirality operator
\bea
\gamma_5 \equiv v^m \Gc_m~.\nn\eea
The Killing spinors have $\gamma_5$-eigenvalue $+1$
\bea
\gamma_5 \xi_I=\xi_I~,\label{helicity_xi}
\eea
as can be shown using the Fierz identities. Thus the bosonic spinor field $q$ is of positive chirality,
 while the fermonic spinor $\psi$ contains both chiralities. Now we introduce an auxiliary  bosonic spinor field ${\cal F}$
  of negative chirality, which satisfies the same reality condition (\ref{reality-cond-spin}) as $\psi$.
   With the auxiliary field, one can obtain an off-shell odd supersymmetry enlarging (\ref{hyper-tran-noaux})
  \bea
  && \delta q^A = iP_+\psi^A~,\nn\\
 && \delta \psi^A =\frac{1}{2r}(t\Gt)_{pq}(\Gc^{pq}q^A)+(\slashed{D}+i\gs) q^A+{\cal F}^A~,\label{susy_hyper_twist}\\
&& \delta {\cal F}^A = -iP_-\slashed{D}\psi^A-\gs P_-\psi^A-\Psi^m(\Gc_m-v_m)q^A~,\nn
\eea
 where $(t\Gt)_{pq}=t_{IJ}\Gt^{IJ}_{pq}$ and we use the projector $P_{\pm}=\frac{1}{2}(1\pm \gamma_5)$ and $P_+ q = q$, $P_- {\cal F}={\cal F}$. Notice that $\gs q^A$ should be understood as $\gs^A_{~B}q^B$ and similarly for the term involving $\Psi_m$.
  The transformations (\ref{susy_hyper_twist})
square off-shell to the following:
\bea
\delta^2\Phi=-\frac{i}{2r}(t\Gt)_{pq}\Gc^{pq}\Phi+\big(iv\cdot D-\gs\big)\Phi~,~~~~\Phi=\{q,\psi,{\cal F}\}~.\label{closure_hyper}
\eea

After a change of variables, one can put the complex (\ref{susy_hyper_twist}) into a nice cohomological form
\bea
 \delta q^A=i\psi^A_+~,&&\gd \psi^A_+=\Big(-\frac1{2r}(t\Gt)_{pq}\Gc^{pq}+v\cdot D+i\gs\Big)q^A~,\nn\\
\gd\psi^A_-=\tilde {\cal F}^A~,&& \gd \tilde{\cal F}^A=\Big(-\frac{i}{2r}(t\Gt_{pq})\Gc^{pq}+iv\cdot D-\gs\Big)\psi^A_-~,\label{coho_hyper_spin}\eea
where $\psi_{\pm}=P_{\pm}\psi$ and $\tilde{\cal F}$ is obtained from ${\cal F}$ by a linear shift (and hence no Jacobian in this change of variables).  The above complex is written in terms of the  fields $q,\psi$ and $\tilde{\cal F}$ which satisfy the reality conditions.
 We can  solve these reality conditions in terms of unconstrained fields as we did in (\ref{rewrite_fermion}). Let $q^A=[q^\alpha,-Cq^*_\beta]^T$, $\psi^A=[\psi^\alpha,-C\psi^*_\beta]^T$ and $\tilde{\cal F}^A=[\tilde{\cal F}^\alpha,-C\tilde{\cal F}^*_\beta]^T$. Now we
  can rewrite the complex (\ref{coho_hyper_spin}) in terms of the new fields and it   looks exactly the same, except for the change of indices $A\to \alpha$
\bea
 \delta q^\alpha=i\psi_+^\alpha~,~~~\delta\psi_+^\alpha=\cdots~.\nn
\eea
One property that we will need for the transformations is that it acts holomorphically, in that it does not mix $q^\alpha,\psi^\alpha,\tilde{\cal F}^\alpha$ with their conjugates.
This point will be important later when we decide over what spaces we compute the determinant of the operator $\delta^2$.

To complete the localization argument, one can add to the action a $\gd$-exact term $\delta\int\Tr V$, with
\bea
V=\frac12(\gd\psi^A)^{\dagger}\psi^A~.\nn
\eea
The bosonic part of $\delta V$ is\footnote{Note the change of gauge indices from $A$ to $\alpha$.}
\bea \gd V=\big(-\frac1{2r} t\Gt_{pq}\Gc^{pq}q+v\cdot Dq\big)_\alpha^{\dagger}\big(-\frac12 t\Gt_{pq}\Gc^{pq}q+v\cdot Dq\big)^\alpha
+(\gs q)^{\dagger}_\alpha(\gs q)^\alpha+\tilde{\cal F}^{\dagger}_\alpha{\cal F}^\alpha~.\nn\eea
In the manipulation above $\gs$ and $\tilde{\cal F}$ are Wick rotated, which is crucial for decoupling the last two terms from the rest. Since $\gd V$ is positive definite the localization locus is given by the following equations
\bea \big(-\frac1{2r} t\Gt_{pq}\Gc^{pq}+v\cdot D\big)q^\alpha=0~,~~~\gs^\alpha_{~\beta} q^\beta=0~,~~~~\tilde{\cal F}^\alpha=0~,\label{locus_hyper}\eea
To further analyze the first condition of (\ref{locus_hyper}), it is convenient
to work with a specific representation of spinors using differential forms, reviewed in Appendix \ref{sec_can_spinc}.
We can represent the fields $q,\;\psi,\;\tilde{\cal F}$ by horizontal anti-holomorphic forms.  To make it more explicit, one can pick a pure spinor, say $\xi_1$, and map the spinors to differential forms in the standard fashion:
\bea
\Phi\to \big\{\xi_1\Phi~,~~~\xi_1\Gc_m\Phi~,~~~\xi_1\Gc_{mn}\Phi\big\}~,~~~\Phi=\{q,\psi,\tilde{\cal F}\}~.\label{conversion}
\eea
That $\xi_1$ satisfies (\ref{pure_spinor}) implies that the above forms are anti-holomorphic. The property (\ref{helicity_xi}) shows that the independent components can be chosen to be horizontal. Chirality consideration tells us that $q$ is mapped to
$\Omega^{0,\textrm{even}}_{H}$, $\tilde{\cal F}$  to $\Omega^{0,\textrm{odd}}_{H}$ and $\psi$ to $\Omega^{0,\textrm{any}}_{H}$.
With an abuse of notations we will use the same letters for the corresponding differential forms.

Now our goal is to rewrite the operator appearing in (\ref{locus_hyper}) in terms of operations on differential forms.
Using (\ref{hor_complex_structure}) and (\ref{gamma_canonical}) one can compute
\bea (t\Gt)_{mn}\Gc^{mn}=2i\big(1-\deg\big)~,\label{nice_relation}\eea
where $\deg$ stands for the degree of the form. As a check one can directly show that $P_-t\Gt_{pq}\Gc^{pq}=0$ from (\ref{duality_curv}) and (\ref{Theta}), and as the $-$ chirality spinor goes to deg 1 forms, we see the agreement with the above relation.

The $+$ chirality subbundle is trivialized by the two sections $\xi_{1,2}$, and we know how $D_m$ acts on $\xi_I$ from the Killing equation. Thus for the $q$ sector, we know the action of $D_m$, and we have
\bea
\big(-\frac{i}{2r}(t\Gt)_{pq}\Gc^{pq}+iv\cdot D\big)q^\alpha=iv\cdot\nabla_A q^\alpha+\frac{1}{r}\big(\frac32-\deg\big)q^\alpha~,\label{temp_2}
\eea
where $\nabla_A$ is the Levi-Civita connection coupled with the gauge field. Notice that we use the same $q$ to denote the forms from the reduction (\ref{conversion}).
In the derivation above we have used the \emph{compatibility of the spin connection with the Levi-connection} (see section 6.1 in the book  \cite{Salamon}), namely, the spin connection induces, through the adjoint action, the Levi-Civita connection on $TS^5$.

To continue, we need to work out how the Levi-Civita connection acts on horizontal differential forms. This is straightforward from the explicit decomposition of the metric (\ref{metric_fibration}), one easily obtains the action of the Levi-Civita connection on horizontal anti-holomorphic forms
\bea
v^m \nabla_{m}= v^m \partial_{m}+\frac12(d\kappa)^{\bar i}_{~\bar j}dx^{\bar j}\iota_{\bar i}~,\nn
\eea
and by relating $d\kappa$ to the complex structure through (\ref{kappa}), (\ref{dkappa}) and (\ref{hor_complex_structure}),
we have
\bea
v^m \nabla_{m}= \L_v-\frac{i}{r}dx^{\bar i}\iota_{\bar i}= \L_v-\frac{i}{r}\deg~,\nn
\eea
 where $\L_v$ is the Lie derivative along $v$ and it is valid to replace $v^m\partial_m$ with $\L_v$ since the forms it acts on are all horizontal. We remark however that
the full actual expression $v^m \nabla_{m}$ has extra terms, it only acquires this simple expression when acting on horizontal anti-holomorphic forms.
Finally putting back the gauge field $A$, the right hand side of (\ref{temp_2}) is written
\bea
\big(iv\cdot\nabla_A q^\alpha+\frac{1}{r}(\frac32-\deg)\big)q^\alpha= i \L_v^A q^\alpha+\frac{3}{2r} q^\alpha~,\nn
\eea
where $\L_v^A$ is defined after (\ref{susy_vect_twist}).
 To obtain the action of $D_m$ on the $-$ chirality sector requires some work, we only remark that the strategy is again to use the Leibniz property and the compatibility between the spin connection and the Levi-Civita conneciton, e.g.\ $\nabla_m(\psi\Gc_n \chi)=(\psi\overleftarrow{D}_m\Gc_n \chi)+(\psi\Gc_nD_m\chi)$.

When this is done, we can rewrite the cohomological complex (\ref{coho_hyper_spin}) in terms of horizontal differential forms $\Omega_H^{0, \bullet}$ as
\bea
&& \delta q^\alpha = i\psi^\alpha_+~,\nn \\
 && \delta \psi^\alpha_+  =  \L_v^A q^\alpha + i\gs q^\alpha - i\frac{3}{2r} q^\alpha~,\nn\\
  && \delta \psi^\alpha_- = \tilde{\cal F}^\alpha~,\label{hypercomplex}\\
  && \delta \tilde{\cal F}^\alpha =  i \L_v^A \psi^\alpha_--\gs \psi^\alpha_-+\frac{3}{2r} \psi^\alpha_-~,\nn\eea
where $\psi_\pm = P_\pm \psi$ and we again remark that  we use the same symbols $q,\psi,{\cal F}$ for the original fields
as well as the differential forms they reduce to.
Here the term $\sigma \Phi$ is understood as $\sigma^a (T_a)^\alpha_{~\beta} \Phi^\beta$ where $T_a$ are the Hermitian generators of
a representation $R$ of $SU(N)$. For the adjoint representation $\sigma \Phi$ is given simply by $[\sigma, \Phi]$ if $\gs$ and $\Phi$ are represented as $N\times N$ matrices.

With the above preparatory work, we can now analyze the zero locus of the operator in (\ref{locus_hyper})
\bea \big(-\frac1{2r} t\Gt_{pq}\Gc^{pq}+v\cdot D\big)= i \L_v + G_{(-i\iota_v A)} + \frac{3}{2r} ~,\nn \eea
where $G_{(-i\iota_v A )}$ is a gauge transformation with parameter $-\iota_v A$.
So far the discussion is valid for any instanton background from the vector multiplet. Next we will specialize to the trivial background
$A=0$.  The zero modes of $q$ will then satisfy
\bea i \L_vq + \frac{3}{2r}q=0~,\nn\eea
any solution must be of the form $q=f(x)e^{3i\gt/(2r)}$, where $\gt\in[0,2\pi r)$ and $v=-\partial_{\theta}$. This however is not a valid solution because of the half integer $3/2$ and the resulting wrong periodicity. Thus one concludes that at the trivial background $A=0$, the localization locus for the hypermultiplet is all bosonic fields being zero.

Next we will calculate the one-loop determinant for the hypermultiplet. The calculation is performed in the same way as we explained in the section for the vector multiplet. As seen from the transformations $\eqref{hypercomplex}$, the fields $q^\alpha$ and $\psi_-^\alpha$ are the analogue of coordinates in a cohomological complex while the fields $\psi_+^\alpha$ and $\tilde{\cal F}^\alpha$ are the corresponding 1-forms with opposite statistics. Therefore the one-loop contribution is given by  the determinant of the operator $\delta^2$ taken on the fields $q^\alpha$ and $\psi_-^\alpha$. As we worked out earlier, when acting on differential forms, the operator $\delta^2$ is given by
\begin{equation}
 \gd^2=i\L_v -i \sigma+\frac{3}{2r}~,\label{hyper-oper-local}
\end{equation}
 where the imaginary "$i$" comes from the integration over imaginary $\sigma$ (this follows from the manipulation in the vector multiplet).
  As also worked out above, the fields in the hypermultiplet are mapped to holomorphic and anti-holomorphic horizontal forms (in spinors this is the same as
   two complex components of $q^\alpha$, four complex components of $\psi^\alpha$ and two complex components of $\tilde{\cal F}^{\alpha}$). Especially, the coordinates $q$ and $\psi_-$ are mapped to
$\Omega^{0,\textrm{even}}_{H}$ and $\Omega^{0,\textrm{odd}}_{H}$, respectively. Therefore the contribution of the hypermultiplet in a representation $R$ is given by the following:
\begin{equation}\label{main-form-det}
Z_{\rm 1-loop}^{\rm hyper}= \prod\limits_\mu \prod\limits_{t}\left( \frac{ t}{r} - \langle i\sigma , \mu\rangle +\frac{3}{2r} \right)^{-(1+\frac{3}{2}t+\frac{1}{2}t^2)}~.
\end{equation}
where the $\mu$'s are the weights of the representation $R$. By our convention $v=-\partial_\theta$ and the range of $\theta$ is  $0\leq\theta<2 \pi r$, which means that the eigenvalues of $\mathcal{L}_v$ is $-\frac{i t}{r}, ~t\in\mathbb{Z}$. As was the case for the vector multiplet, the modes labelled by $t$ is a section of the bundle ${\cal O}(t)$ over $\BB{C}P^2$, and the expression in the exponent in $\eqref{main-form-det}$ is determined by using the Atiyah-Singer index theorem for the Dolbeault complex twisted by ${\cal O}(t)$. Again, for more technical details of these types of calculations, we refer to \cite{KallenZabzine12} and references therein.

In order to analyze the expression (\ref{main-form-det})  let us study the following infinite product
\be
 h(x) = \prod_{t\in\BB{Z}} \left (t - ix + \frac{3}{2} \right )^{-(1+ \frac{3}{2} t + \frac{1}{2} t^2)}~,~~~x\in \BB{R}
\ee
 which has a number of curious properties. If one first flips $t$ to $-t$ and then shifts $t \rightarrow t+3$, then one gets
 the following relation
   \be
    h(x) = h(-x) \prod_t (-1)^{-(1+ \frac{3}{2} t + \frac{1}{2} t^2)}~.\label{symmetry_h}
   \ee
    Using $\zeta$-function regularization we arrive at the property that $h(x) = h(-x)$. This is the same as saying that
     $h(x)$ is a real and even function\footnote{It should be remarked that as the original product is ill-defined without regularization, the shift $t\to t+3$ in the previous manipulation is at best formal, one should look for a more rigorous argument. However, once we arrive at (\ref{def-h-firstone}) we can check explicitly these properties using the inversion formulae of polylogs, see equation $\eqref{magic-ell-f}$.} of $x$. Next we can rewrite $h(x)$ as follows
\begin{equation}
\begin{split}
h(x) = \prod\limits_{t} \left( (t+1)  -ix +\frac{1}{2} \right)^{-\frac{1}{2}(1+t + (1+t)^2)}= \prod\limits_{t}\left( t  -ix  +\frac{1}{2} \right)^{-\frac{1}{2}(t + t^2)} ~.
\end{split}
\end{equation}
Using the definitions (\ref{Jaff-func-exp}) and (\ref{our-func-exp}) we can write $h(x)$ as
\be
 h(x) = e^{\zeta'(-2)} e^{\frac{1}{2} \ell \left (ix - \frac{1}{2} \right )} e^{-\frac{1}{2} f \left (\frac{1}{2} - ix \right )}~,\label{def-h-firstone}
\ee
 where we have used $\zeta$-function regularization.  The asymptotic behaviour of $h$ is dominated by $f$
\bea \lim_{x\to\pm\infty} f(ix)\sim -\frac{\pi}{3}|x|^3~.\nn\eea

 Using the symmetry property $h(x) = h(-x)$ we rewrite
 \be
 && h(x) = e^{\zeta'(-2)} e^{\frac{1}{4} \ell \left (ix - \frac{1}{2} \right ) + \frac{1}{4} \ell \left (-ix - \frac{1}{2} \right )}
   e^{-\frac{1}{4} f \left (\frac{1}{2} - ix \right ) - \frac{1}{4} f \left (\frac{1}{2} + ix \right )}  \nn \\
   && = e^{\zeta'(-2)} \left ( 2 \cos (\pi i x) \right )^{\frac{1}{4}}  e^{-\frac{1}{4} f \left (\frac{1}{2} - ix \right ) - \frac{1}{4} f \left (\frac{1}{2} + ix \right )}
 \ee
  Using this expression for $h(x)$ the contribution (\ref{main-form-det}) can be written as follows:
 \be
 Z_{\rm 1-loop}^{\rm hyper}= \prod\limits_\mu e^{\zeta'(-2)} \left ( 2 \cos (\pi \langle i\sigma {r}, \mu \rangle ) \right )^{\frac{1}{4}}  e^{-\frac{1}{4} f \left (\frac{1}{2} - \langle i\sigma {r}, \mu \rangle  \right ) - \frac{1}{4}
  f \left (\frac{1}{2} + \langle i\sigma {r}, \mu \rangle  \right )}~.
\ee
  Putting everything together the final result for the $N=1$ vector multiplet coupled to a hypermultiplet in a representation $R$
   is given by the following matrix model:
\be
Z&=&\int\limits_{\rm Cartan} [d\phi]~e^{-  \frac{ 8\pi^3 r}{g_{YM}^2}  \text{Tr}(\phi^2)} {\rm det}_{\rm Ad}\left (  \sin ( i\pi  \phi) e^{ \frac{1}{2} f(i \phi )}  \right ) \nn \\
&&\times~  {\rm det}_{R} \left ( \left ( \cos ( i\pi \phi )\right )^{\frac{1}{4}} e^{-\frac{1}{4} f \left (\frac{1}{2} -  i\phi \right ) - \frac{1}{4} f \left (\frac{1}{2} +  i\phi \right )} \right ) ~,\label{vector1loop2}
\ee
  where the  dimensionless combination $\phi = r \sigma$ is used.  In this expression we ignore possible irrelevant overall
   numerical factors. For the case of the hypermultiplet in the adjoint representation, the answer can
    be simplified a bit to
\be
 \int\limits_{\rm Cartan} [d\phi]~e^{-  \frac{8\pi^3 r}{g_{YM}^2}  \text{Tr}(\phi^2)} \prod_{\beta >0} (\sin (\pi \langle
 \beta, i \phi \rangle))^2 \sqrt{\cos(\pi \langle \beta, i\phi \rangle)} e^{f(\langle \beta, i \phi \rangle) - \frac{1}{2}
   f\left (\frac{1}{2} -\langle \beta, i \phi \rangle \right )
 - \frac{1}{2} f\left ( \frac{1}{2} + \langle \beta, i \phi \rangle \right )}~.\nn
\ee
  Let us make some remarks on these matrix models.
 One can  notice the relative sign between the $f$-function contributions  in (\ref{vector1loop2}), and using the asymptotic behavior of $f$, one realizes that if the matter is in a representation of large enough dimension, the matrix model potential flips sign at infinity and becomes unstable. In the case the matter is in the adjoint, its leading $|x|^3$ behavior cancels  that of the vector multiplet, and the subleading behavior is
\be
\lim_{x\to\pm\infty} \left [  f(ix) -\frac{1}{2} f \left ( \frac{1}{2} -ix \right ) - \frac{1}{2} f \left ( \frac{1}{2} + ix \right ) \right ] \sim -\frac{\pi}{4}|x|~,
\ee
 which is tamed by the overall Gaussian damping. Thus one adjoint hypermultiplet seems to be the marginal matter content,
  beyond which it seems that the matrix model makes little sense.

This final matrix model corresponds to the $N=2$ model, at least as far as the field content is concerned; one vector plus one hypermultiplet in the adjoint.
  In the flat space limit, one $N=1$ vector multiplet plus one hyper-multiplet in the adjoint representation gives an $N=2$ model. We do not claim that the same $N=2$ susy theory can be put on $S^5$. But as one takes the limit $r\to \infty$, our model is identical to that of the $N=2$ model. Thus we expect our result to be certain deformation of the flat $N=2$ model, which allows us to compute the partition function of the non-renormalizabile $N=2$ model. Further discussion on the possibility of understanding our result in the light of the 6D (2,0) model is given in  section \ref{summary}.

\subsection{Comment on phases of determinants}

So far we have computed  the absolute value of the one-loop determinants.  Typically in odd dimensions additional phases
 may appear when one calculates the determinant of a Hermitian operator with an unbounded spectrum. The phase originates from
  the mismatch between positive and negative eigenvalues and it requires an additional regularization. The typical calculation
  of an additional phase in 3D Chern-Simons theory in the localization framework can be found in  \cite{Beasley:2005vf} (see also \cite{Kallen11} for a discussion in a
   similar context), where it gives rise to the famous shift of the Chern-Simons level.  Naively generalizing these arguments the phase has been calculated in 5D topological Yang-Mills theory
     \cite{KallenZabzine12} (also see \cite{Losev:1995cr} for a related early discussion),
  giving the following additional term in matrix model
  \begin{equation}
e^{i\frac{3\pi }{2}\sum\limits_{\beta>0}{\langle \beta, \phi \rangle^2}}~.
\end{equation}
 However in the present work we are forced to work in conventions where the appropriate operators (\ref{oper-vect-1loop})
 and (\ref{hyper-oper-local})
  are not Hermitian. Furthermore, for the hypermultiplet, the symmetry property such as (\ref{symmetry_h}) shows that the phase is zero formally. It is possible that some subtlety has been overlooked in our formal manipulation, yet on the other hand we do not have an
   alternative derivation or physical indication that the phase should be there.

\section{Summary}
\label{summary}

In this section we want to summarize our results and point out some open problems.

 By using the localization technique we were able to calculate the full perturbative partition function for the deformed
  $N=1$ 5D  supersymmetric Yang-Mills theory with matter. The result is given in terms of a matrix model which, as far as
   we can see, is well-defined. In general one needs to analyze the non-perturbative corrections for this partition function
    which comes from instantons on $\mathbb{C}P^2$. However, for the case of an abelian gauge group the contact instantons
     on $S^5$ corresponds to the flat connection.

 The main puzzle to us comes from the fact that 5D Yang-Mills is perturbatively non-renormalizable. By putting this theory
  on the sphere we do not modify the UV behavior of the theory. At the same time we are able to produce a well-defined answer
   and respect all rules of the game. However we have to keep in mind that localization typically comes with a whole package,
   including things such as $\zeta$-function regularization and analytical continuation. At the moment we do not understand
    the relation of our analysis to the discussion of UV properties of 5D Yang-Mills theory.

 Let us comment on a possible interpretation of our result. Once we define the dimensionless matrices $\phi$ our matrix model
  depends only on the ratio $\frac{r}{g_{YM}^2}$ in front of the Gaussian term. If we naively would like to take decompactification limit
   $r\rightarrow \infty$, then the matrix model will collapse. However we can send $r$ to infinity together with $g_{YM}^2$ to infinity
    while keeping their ratio fixed. We can therefore keep our matrix model result if we go to the flat space limit and strong coupling simultaneously.
 Remember that the coupling of the 5D Yang-Mills theory is related to the magic 6D theory as follows:
\bea
\frac{1}{g_{YM}^2} = \frac{1}{r_6}~,
\eea
 where $r_6$ is the radius of a circle added to the five manifold to get the 6D theory.  From this point of view our matrix model depends on
  \bea
  \frac{r}{g_{YM}^2} = \frac{r}{r_6}
  \eea
 and if  we keep this ratio fixed then  the answer is still meaningful when we send $r\rightarrow \infty$ and $r_6 \rightarrow \infty$.
  This is again a formal indication that our matrix model can eventually be related to the 6D theory. The main check for this would be
   to study the large $N$-limit and to reproduce the famous $N^3$ behavior. Unfortunately the present matrix model is quite complicated
    and this is not a straightforward problem. We hope to say something meaningful about this problem in the future.

\bigskip\bigskip

\noindent{\bf\Large Acknowledgement}:
\bigskip

\noindent  We thank  Joseph Minahan, Vasily Pestun and Konstantin Zarembo for useful discussions
 on this and related subjects.   M.Z. thanks INFN Sezione di Firenze, Universit\`a
di Firenze   where part of this work was carried out.  The research of MZ is supported by VR-grant 621-2011-5079.
\\
\\
\noindent{\bf\large Note added in proof} 
\\
While this paper was under review, it was shown in \cite{Kallen:2012zn} that for the model with one vector multiplet and one hypermultiplet in the adjoint representation, the free energy does indeed scale as $N^3$ at large Õt Hooft coupling

\appendix
\section{Convention for spinors and Fierz identities}
\label{A-spinors}

We follow the convention for spinors from \cite{HosomichiSeongTerashima}. The gamma matrices satisfy the Clifford algebra
\bea
\{\Gc^{\tt a},\Gc^{\tt b}\}=2\delta^{\tt ab}~,\nn
\eea
and the charge conjugation matrix satisfies
\bea
C^{-1}(\Gc^{\tt a})^TC=\Gc^{\tt a}~,~~~~C^T=-C,~C^*=C~.\nn
\eea
The spinor bi-linears are formed using $C$,
\bea
\psi^TC\chi\stackrel{\textrm{abbreviate}}{\longrightarrow}\psi\chi~,\label{spinor_bi_linear}
\eea
though throughout the paper, these bi-linears are abbreviated as $(\psi\chi)$, following the notation of \cite{HosomichiSeongTerashima}. Due to the symmetry property of $C$, one has
\bea
(\psi\chi)=-(\chi\psi),~~(\psi\Gc^{\tt a}\chi)=-(\chi\Gc^{\tt a}\psi)~,~~(\psi\Gc^{\tt ab}\chi)=(\chi\Gc^{\tt ab}\psi)~,\nn
\eea
where $\Gc^{\tt a_1\cdots a_n}=(1/n!)\Gc^{[\tt a_1}\cdots \Gc^{\tt a_n]}$ and all spinors appearing above are \emph{bosonic} (even). The product of three or more gamma matrices can be reduced
\bea
 \Gc^{\tt abc}e^{\tt abcde}=-6\Gc^{\tt de}~,\label{duality_flat}
 \eea
where $e^{\tt a...b}$ is the Levi-Civita symbol $e^{12345}=1$.

On a curved manifold, one defines the gamma matrices by means of the veilbeins, i.e.\ a set of mutually orthogonal (local) sections of the tangent bundle
\bea E^{\tt a}\in \Gc(TM)~,~~~\bra E^{\tt a},E^{\tt b}\ket=\delta^{\tt ab}~,\nn
\eea
where $\bra-,-\ket$ is the pairing using the metric $g$.
 The gamma matrices are defined as
\bea
\Gc^m=E^{m\tt a}\Gc^{\tt a}~,~~~\Gc_m=g_{mn}\Gc^n,\nn
\eea
and the duality (\ref{duality_flat}) turns into
\bea
\frac{1}{3!}g^{1/2}\Gc_{mnp}\epsilon^{mnp}_{~~~~qr}=-\Gc_{qr}~.\label{duality_curv}
\eea

The following identity is called the Fierz relation ($\zeta,\eta,\psi$ are bosonic spinors)
\bea \zeta(\eta\psi)=\frac14 \psi(\eta\zeta)+\frac14 \Gc^m\psi(\eta\Gc_m\zeta)-\frac18 \Gc^{mn} \psi(\eta\Gc_{mn}\zeta)~.\label{Fierz_2}\eea
From this one can derive another useful identity
\bea
 \Gc_m\chi(\eta\Gc^m\psi)+\chi(\eta\psi)=2\psi(\eta\chi)-2\eta(\psi\chi)~.\label{most_useful}
 \eea

As an example of a typical manipulation used in the paper let us  derive the $v^2=1$ relation, which is also derived in \cite{HosomichiSeongTerashima}. Recall that
 $v^m$ is defined as
\bea
v^m=\epsilon^{IJ}\xi_I\Gc^m\xi_J=\xi_I\Gc^m\xi^I~,\nn
\eea
where $\xi_I$ are defined in subsection \ref{sec_cov} with normalization given by (\ref{normalize}).
 We fix the following convention for the manipulations with the $SU(2)_R$ indices:
 \bea
   \xi^I=\epsilon^{IJ}\xi_J~,~~~~ \xi_I=\epsilon_{IJ}\xi^J~,~~~~ \epsilon^{IK}\epsilon_{KJ}=\delta^I_J~,~~~~ \epsilon^{12}=-\epsilon_{12}=1~.\nn
   \eea
Then one has
\bea v^2=(\xi_I\Gc^m\xi^I)(\xi_J\Gc_m\xi^J)=-(\xi_I\xi^I)(\xi_J\xi^J)+2(\xi_I\xi^J)(\xi_J\xi^I)-2(\xi_I\xi_J)(\xi^J\xi^I)=(\xi_I\xi^I)^2=1~.\nn\eea
 Throughout the paper the Fierz relations are sometimes used without mentioning it explicitly.

\section{Five-manifolds admitting two Killing spinors}
\label{app-Killing}
In this section $\nabla$ is reserved for the Levi-Civita connection while $D$ denotes the spin connection.

Consider a five-manifold admitting a pair of normalized Killing spinors
\bea
     \xi_I\xi_J=-\frac{1}{2}\epsilon_{IJ}~,~~~~~~D_m\xi_I =\frac{1}{r} t_I^{~J}\Gc_m\xi_J~.
\eea
 We are interested in studying the geometrical consequence of these equations.
 Using $\xi_I$ we can define a vector field
\bea
v^m=\xi_I\Gc^m\xi^I~,~~~g_{mn}v^mv^n=1~,\label{Reeb}
\eea
where the normalization can be derived using the Fierz relations listed in Appendix \ref{A-spinors}.
Next define a 1-form $\kappa$ as follows:
\bea
\kappa_m =g_{mn}v^n~.\label{kappa}
\eea
Calculate its covariant derivative by using the Killing equation:
\bea \nabla_m \kappa_n=-\frac{2}{r} t_{IJ}\Theta^{IJ}_{mn}~,~~~~\Theta^{IJ}_{mn}=\xi^I\Gc_{mn}\xi^J~.\label{dkappa}
\eea
We will use the object $\Theta$ in many places in this paper. Since $\Theta$ is antisymmetric in $m,n$, one
 deduces $\L_v g_{mn}=\nabla_{(m}\kappa_{n)}=0$.
Moreover we have
\bea
 \iota_v\Theta^{IJ}=0~,~~~ \iota_v*\Theta^{IJ}=-\Theta^{IJ}~,~~~
\sqrt{g}~\Theta_{mn}^{IJ}\Theta^{KL}_{pq}\epsilon^{mnpq}_{~~~~~\;r}=2\epsilon^{L(I}\epsilon^{J)K}\kappa_r~.\label{Theta}\eea
The second relation shows $\iota_v*d\kappa=-d\kappa$ and the last relation implies $\kappa\wedge(d\kappa)^2\neq0$ and thus $\kappa$ is a contact form  with $v$  being the corresponding
 Reeb vector. Thus we conclude that the manifold is equipped with a $K$-contact structure (a contact structure with a compatible metric
  such that the Reeb vector field is a Killing vector for this metric).  As an example of this situation we can assume that
  the Reeb vector corresponds to a $U(1)$ action on the underlying five-manifold.

Having a nowhere vanishing vector field $v$, the structure group of $M_5$ can be reduced to $SO(4)$. Next we
  define a complex structure in the plane orthogonal to $v$ (it is called the contact plane in the context of contact geometry) by
\bea
 J_m^{~\;n}= 2 t_{IJ}\Gt^{IJ}_{mp}g^{pn}~,~~~J_m^{~\;p}J_p^{~n}=-(\delta_m^n-\kappa_mv^n)\label{hor_complex_structure}
 \eea
which implies that the structure group is reduced to $U(2,J)$.
 This is a general feature for contact manifolds, see for example Chapter 8 in the book \cite{Geiges}.
  Using the Fierz relations one can also show that the complex structure satisfies
\bea
\nabla_pJ_m^{~\;n}=\frac{1}{r} \left ( g_{pm}v^n-\delta^n_p\kappa_m \right )~.\label{kahler}
\eea
It can be shown that this complex structure induces an almost complex structure on the base $M_4$, and using the above relation one can conclude that $M_4$ is K\"ahler. But we will not spell out the detailed argument here.

Furthermore we have
\bea J_p^{~q}\Gc_q\xi_K=- 2 t_K^{~\;I}(\Gc_p\xi_I)+ 2 t_K^{~\;I}v_p\xi_I~,\nn\eea
which implies
\bea \big(P(1+iJ)\big)_p^{~q}\Gc_q\xi_2=0~,~~~~\big(P(1-iJ)\big)_p^{~q}\Gc_q\xi_1=0~,~~~~\textrm{where}~~P_m^n=\delta_m^n-\kappa_m v^n~.\label{pure_spinor}\eea
The two spinors $\xi_I$ are \emph{pure spinors} for $J$ and $-J$ respectively (see (\ref{pure_spinor_def}) for a definition of pure spinors), this
 will play a role when one maps spinors to differential forms. One can derive more conditions on the K\"ahler base, such as the relation between the canonical bundle and the K\"ahler class, but we will not digress too much here. For the concrete case of $S^5$ viewed as a $S^1$ fibration over $\mathbb{C}P^2$ an explicit construction of Killing spinors is given in Appendix \ref{A-KIll-spin}.

\section{The canonical Spin${}^c$ representation}
\label{sec_can_spinc}

Consider a complex vector bundle $V\to M$ with complex structure $J$ and a metric which is Hermitian with respect to $J$.
 A \spinc-structure is a lifting of the structure group $SO(V)$ to
\bea
\textrm{Spin}^c(V)=\textrm{Spin}(V)\times_{\BB{Z}_2}U(1)~.\label{naive}
\eea
We will not extensively review \spinc-structures here (see the book \cite{Salamon}, section 5.1 for a nice treatment), but the rough idea is that when one tries to lift $SO(V)$ to $\textrm{Spin}(V)$ one faces a $\BB{Z}_2$ worth of choice as the latter group is a two sheeted cover of the former. The local choices will have some incompatibilities globally, which is measured by the second Stiefel-Whitney class $w_2(V)$. In the case $w_2(V)$ has an integral lift, one can construct a line bundle (called the \emph{characteristic line bundle} of the \spinc-structure) whose 'square root' has the same global incompatibility as above. The intuitive picture (\ref{naive}) exactly reflects the fact that neither factor on the right hand side is a bona fide bundle, but together the incompatibility cancel and they make a globally defined lifting of $SO(V)$.

One can construct an associated \spinc-bundle as follows. Consider the vector bundle
\bea
W=\wedge^{0,\sbullet}V^*~.\label{canonical_spinc}
\eea
Choose a basis of $V^*$ as: $e^{i}\in \ker (1+iJ)$ and $e^{\bar i}\in \ker (1-iJ)$, then one can define the Clifford action as
\bea
X\cdot \psi=\sqrt2 \big(X^ig_{i\bar i}e^{\bar i}\wedge+X^{\bar i}\iota_{e^{\bar i}}\big)\psi~,~~~X\in V~,~~~\psi\in W~.\label{canonical_clifford}\eea
This is the canonical \spinc-structure on a complex vector bundle (see section 5.3 in \cite{Salamon}).
To construct a \spinc-connection, it is useful to keep in mind the intuitive idea (\ref{naive}), the connection consist of two parts
\bea
D=d+\omega+\frac{i}{2}A~,\label{spin_conn_2}
\eea
the first part $\omega$ is the lift of $\FR{so}(V)$ to $\FR{spin}(V)$ given by the standard formula
\bea
M^{\tt ab}\to \frac14 M^{\tt ab}\Gc^{\tt ab}~,~~~~M^{\tt ab}\in\FR{so}(V)~.\label{lift_so_spin}
\eea
The second part is {\emph half} of the connection $A$ of the characteristic line bundle.

A \emph{pure spinor} of a \spinc-representation is a subspace of $E_J\subset W$
 \bea
  E_J=\{\psi\in W| (1-iJ)X\cdot \psi=0,~\forall X\in V\}~.\label{pure_spinor_def}
  \eea
In other words, the pure spinor is annihilated by $\Gc^{i}$ under the Clifford multiplication. In the case of a canonical representation (\ref{canonical_spinc}),  the pure spinor is the subbundle $\wedge^{0,0}V^*$. Clearly, this subbundle does not exist unless the structure group \spinc~ is reduced to
\bea
U^c(V,J)=\{x\in\textrm{Spin}^c(V)|ad(x)\in U(V,J)\}~,\nn\eea

This construction enters our computation as follows. Assume that the orbit of the Reeb vector field is closed everywhere, which means our five-manifold is the total space of a $U(1)$ bundle over a 4-manifold
\bea \begin{array}{l}
       M_5 \leftarrow  U(1) \\
       \;\downarrow\pi    \\
       M_4   \end{array}~.\nn\eea
Due to the relation $\L_v g_{mn}=0$ one can assume that the metric is written as
\bea g=(d\theta+\mathscr{A})\otimes(d\theta+\mathscr{A})+g^{\textrm{base}}~,\label{metric_fibration}\eea
where $\mathscr{A}$ is the connection of the principle $U(1)$-bundle and $\gt\in[0,2\pi)$ is the coordinate of the fibre. Note that the combination
\bea -\kappa=d\theta+\mathscr{A}\nn\eea
is independent of the trivialization and is in fact the contact form.

From now on, we will use $\{x^{\mu}\}$ or $\{x^i,\;x^{\bar i}\}$ for the coordinates of $M_4$ and $\{x^m\}=\{x^{\mu},\gt\}$ as that of the five-manifold. Since one can choose the transition function of the $U(1)$-bundle to be independent of $\gt$, the 1-forms $\{dx^{\mu}\}$ are transformed to $\{dx^{\mu}\}$, but $d\gt$ will mix with $\{dx^{\mu}\}$ under a change of trivialization. This consideration shows that the canonical \spinc-bundle introduced in (\ref{canonical_spinc})
  can be concretely written as (taking $V^*=\pi^*T^*M_4$)
\bea
W=\pi^*\Omega^{0,\sbullet}(M_4)\nn\eea
with the explicit Clifford action
\bea\
\Gc_{\mu}=\cA_{\mu}(-1)^{\deg+1}+\sqrt2g^{\textrm{base}}_{\mu\bar i}dx^{\bar i}+\sqrt2\delta_{\mu}^{\bar i}\iota_{{\bar i}}~,~~~~~\Gc_{\gt}=(-1)^{\deg+1}~.\label{gamma_canonical}\eea
The first term in $\Gc_{\mu}$ is perhaps not so clear at first sight, but it arises because the metric $g$ of $M_5$ has a component $g_{\mu\gt}=\cA_{\mu}$. Other than this point, the rest is clear from (\ref{canonical_clifford}).

\section{Solving for Killing spinors on $S^5$}
\label{A-KIll-spin}

This section is rather independent of the main text and can be skipped with no harm to the completeness of the paper.
 In the main text we use  the fact that there is a pair of normalized Killing spinors on $S^5$ (this can be established by using B\"ar's cone construction \cite{cone_construction} and the formula is also given in \cite{HosomichiSeongTerashima}) and from this we deduced  the spin connection acting on anti-holomorphic horizontal
 forms. In this appendix, we will start from the metric
  (\ref{metric_fibration}) and obtain from scratch the spin connection and the Killing spinors in terms of horizontal anti-holomorphic forms, as an independent confirmation of the results used in the main text. With the explicit Killing spinors, relations like (\ref{conversion}) and (\ref{nice_relation}) become quite clear.

Let $[z^1,z^2,z^3]$ be homogeneous coordinates of $\BB{C}P^2$. In the patch $z^3\neq0$, we can use the inhomogeneous coordinates $x^1=z^1/z^3,~x^2=z^2/z^3$. Then the connection $\cA$ for the $U(1)$-fibration $S^5\to\BB{C}P^2$ is (see for example around equation (2.81) in \cite{HosomichiSeongTerashima} for the derivation)
\bea
&&\cA_i=-\frac{i}{2}\partial_iK~,~~~\cA_{\bar i}=\frac{i}{2}\partial_{\bar i}K~,~~~K=\log(1+|x^1|^2+|x^2|^2)~,~~~g_{i\bar i}=\frac12\partial_{i}\partial_{\bar i}K.\nn
\eea
As it has been discussed in a previous Appendix, a K\"ahler manifold possesses a canonical \spinc-structure (\ref{canonical_spinc})
 $W=\Omega^{0,\sbullet}(\BB{C}P^2)$, with the Clifford action defined therein with $e^{\bar i}=dx^{\bar i}$. In particular, the characteristic line bundle of the \spinc-structure is the anti-canonical bundle, which for $\BB{C}P^2$ is ${\cal O}(3)$. One can follow the discussion around  (\ref{spin_conn_2})
   to construct the \spinc-connection, and it is exactly the Levi-Civita connection acting on $W$ (which is in general not true except for K\"ahler manifolds). The details can be found in lemmas 6.10 and 6.11 from the book \cite{Salamon}. For the record, the canonical \spinc-structure is the unique one on a (non CY) K\"ahler manifold to admit a parallel spinor \cite{AndreiMoroianu}, which is just $1\in\Go^{0,0}$.

We construct the spin connection on $S^5$ by the same path as above. One can choose a vielbein by lifting the vierbein of $\BB{C}P^2$ using the metric (\ref{metric_fibration}). One then computes the Levi-Civita connection and rewrites it in the vielbein basis, which is of course $\FR{so}(5)$ valued. This then can be lifted to be $\FR{spin}(5)$-valued according to (\ref{lift_so_spin}). This gives us the first part of the right hand side of (\ref{spin_conn_2}). For the second part, we should choose $A=-3\cA$, where the factor $3$ is from the fact that the anti-canonical bundle $\bar K_{\BB{C}P^2}$ is ${\cal O}(3)$. But we notice that ${\cal O}(3)$ is pulled back to a trivial line bundle on $S^5$, for which the most immediate manifestation is that we can write the curvature $\mathscr{F}=d\cA=d(d\gt+\cA)$ on $S^5$ and $d\gt+\cA=-\kappa$ is a global 1-form. We can use this freedom to shift
\bea A\Rightarrow A-3\kappa=3d\gt~.\nn\eea
To summarize this extremely sketchy construction, we record explicitly the action of the spin-connection on $W=\pi^*\Go^{0,\sbullet}(\BB{C}P^2)$
\bea D_{\mu}-\cA_{\mu}D_{\gt}&=&(\partial_{\mu}-\cA_{\mu}\partial_{\gt})-\Gamma_{\mu\bar j}^{\bar i}dx^{\bar j}\iota_{dx^{\bar i}}
-\frac1{2\sqrt 2}\big(2ig^{\textrm{base}}_{\mu\bar i}dx^{\bar i}-2i\delta_{\mu}^{\bar i}\iota_{dx^{\bar i}}\big)(-1)^{\deg}~,\nn\\
D_{\gt}&=&\partial_{\gt}+i\deg+\frac{i}{2}~,\nn\eea
where $\Gamma^{\bar i}_{\mu\bar j}$ and $g^{\textrm{base}}$ are the Levi-Civita connection and metric of $\BB{C}P^2$.

We seek solutions of the Killing equation among $\pi^*\Go^{0,\textrm{even}}(\BB{C}P^2)$, the Killing equation reads (using (\ref{gamma_canonical})  for the gamma matrices)
\bea
\deg\psi=0~:&&  (\partial_{\mu}-\cA_{\mu}\partial_{\gt})\psi
-\frac{i}{\sqrt 2}g^{\textrm{base}}_{\mu\bar i}dx^{\bar i}\psi=\sqrt{2}cg^{\textrm{base}}_{\mu\bar i}dx^{\bar i}\psi~,\nn\\
&& (\partial_{\gt}+\frac{i}{2})\psi=-c\psi~,\nn\\
\deg\psi=2~:&&  (\partial_{\mu}-\cA_{\mu}\partial_{\gt})\psi-6i\delta_{\mu}^{\bar i}\cA_{\bar i}\psi
+\frac{i}{\sqrt 2}\delta_{\mu}^{\bar i}\iota_{\bar i}\psi=\sqrt{2}c\delta_{\mu}^{\bar i}\iota_{\bar i}\psi~,\nn\\
&&(\partial_{\gt}+\frac{5i}{2})\psi=-c\psi~.\nn\eea
For the first case, it is clear that the solution is $c=-i/2$ and $\psi=1$. For $\deg\psi=2$, we have $c=i/2$ and $\psi=-\rho$ with
\bea
\rho=\frac{1}{2\pi^{3/2}}e^{-3i\gt}(1+|x^1|^2+|x^2|^2)^{-3/2}d x^{\bar 1}\wedge dx^{\bar 2}~.\label{rho_sample}
\eea
This is the expression in the patch $z_3\neq0$, in the other patches, the expression is obtained by cyclically rotating the labels $1,2,3$. One can check explicitly that the expression of $\rho$ defined patchwise is actually global: the transformation of $\gt$ cancels the transformation of the rest. We also remark that, the existence of the non-vanishing global section $\rho$ allows one to write the charge conjugation operator on $S^5$.

\section{Expansions for special functions}
\label{A-special-functions}

In this Appendix we present the infinite product expansions for the special functions used in the paper.
 We need the following infinite products:
\be
 \frac{\sin (\pi y)}{\pi y} = \prod_{t=1}^{\infty}  \left ( 1 - \frac{y^2}{t^2} \right)~,\label{infin-exp-sin}
\ee
\be
 e^{\ell (y)}   = \prod_{t=1}^{\infty} \left ( \frac{y+ t}{y - t} \right )^t = \prod_{t\neq 0} \left (1 + \frac{y}{t} \right )^t~,\label{Jaff-func-exp}
\ee
\be
 e^{f(y)} =  \prod_{t=1}^{\infty} \left (1  - \frac{y^2}{t^2} \right )^{t^2}~.\label{our-func-exp}
\ee
 The formula (\ref{infin-exp-sin}) is the standard representation of sin as an infinite product.
  The formula (\ref{Jaff-func-exp})  defines the function $\ell(y)$ which appeared previously in Jafferis's work \cite{Jafferis:2010un}
   on localization in 3D theories with matter.  The formula (\ref{our-func-exp}) defines the function $f(y)$ which appeared
    previously in \cite{KallenZabzine12} in the study of 5D topological Yang-Mills theory.

  Let us review the explicit expressions for $\ell(y)$, $f(y)$ and some of their properties.
   The function $\ell(y)$ satisfies the following equation
   \be
   \frac{d \ell}{d y} =\sum_{t=1}^\infty \left ( \frac{t}{y+ t} + \frac{t}{t-y} \right ) = 2 \sum_{t=1}^\infty \frac{t^2}{t^2 - y^2}=
    2 \sum_{t=1}^\infty 1 + 2y^2 \sum_{t=1}^\infty \frac{1}{t^2 - y^2}~.
   \ee
     Using $\zeta$-function regularization and the expansion of $\cot$ we arrive at the equation
     \be
      \frac{d \ell}{d y} = - \pi y \cot (\pi y)~.
     \ee
      Upon integration we arrive at the following function
\be
\ell(y)= -y \cdot \ln{(1-e^{2 i \pi y})}+\frac{i\pi y^2}{2}+\frac{i}{2\pi}\text{Li}_2(e^{2 i\pi y})- \frac{i\pi}{12}~. \label{def-ell-appen}
\ee
Our notation is that $\ln$ denotes the logarithm in its principle branch $-\pi<\Im\ln z\leq\pi$. All polylogs $\text{Li}_s(z)$ appearing are also in their principle branch, where the only branch point is $z=1$.

We will need the following inversion formula that relates polylogs to the Bernoulli polynomials:
\bea &&i^{-2}Li_2(e^{2\pi iy})+i^{2}Li_2(e^{-2\pi iy})=2\pi^2\big(-\frac16+y-y^2\big)~,\nn\\
&&i^{-3}Li_3(e^{2\pi iy})+i^{3}Li_3(e^{-2\pi iy})=-\frac43\pi^3\big(y^3-\frac32y^2+\frac12y\big)~, \label{dialog-prop}
\eea
where the domain of validity is $\{y|0\leq\Re{y}<1,\Im{y}\geq0\}\cup\{y|0<\Re{y}\leq1,\Im{y}<0\}$. We remark that it is the restriction on $y$ that caused the apparent mismatch of symmetry properties between the left and right hand sides of the above formula.
Using (\ref{dialog-prop}) one can derive the following important identity
\be
\ell \left (y+\frac{1}{2}\right )+\ell \left (-y+\frac{1}{2} \right )= - \ln(2\cos{(\pi y)})~.
\ee
We will only need this formula for $y$ purely imaginary.

Next let us review properties of $f(y)$. The function $f(y)$ satisfies the equation
\be
 \frac{d f}{dy} = \sum_{t=1}^\infty \left ( \frac{t^2}{t+y} - \frac{t^2}{t-y} \right ) = \sum_{t=1}^\infty \frac{2y t}{y^2 - t^2} =
  - 2y \sum_{t=1}^\infty 1 + 2 y^3 \sum_{t=1}^\infty \frac{1}{y^2 - t^2}~.
\ee
 Using $\zeta$-function regularization and the expansion of $\cot$ we get
 \be
  \frac{ d f}{d y} = \pi  y^2 \cot (\pi y)~.
 \ee
 Upon integration we get the following function
 \be
 f (y)=\frac{i \pi y^3}{3}+y^2 \ln{(1-e^{-2 \pi i y})}+ \frac{iy}{\pi} \text{Li}_2(e^{-2  \pi iy})+\frac{1}{2\pi^2}\text{Li}_3(e^{-2 \pi i y})-
 \frac{\zeta(3)}{2\pi^2}~.\label{definition-f-func}
\ee
We also need a few useful relations between the functions $\ell(y)$ and $f(y)$. Using the definitions
(\ref{def-ell-appen}) and (\ref{definition-f-func}) as well as the inversion formulae for the polylogs
we can show that the following identities hold, assuming $\Im y\neq0$ and $\Re y\in[0,1)$ (which is the range of $y$ we shall need)
\be
 \ell(-y) = -\ell(y)~,~~~f(-y)= f(y)~.
\ee
We remark that at first sight, $\ell(y)$ and $f(y)$ have branch cuts for $\Im y<0$ and $\Im y>0$ respectively, yet a close inspection shows that the branch behaviour cancel. One can also show the following
\begin{equation}\label{magic-ell-f}
 \ell\big(y -\frac{1}{2}\big) - f\big(-y +\frac{1}{2}\big)=\ell\big(-y -\frac{1}{2}\big) - f\big(y +\frac{1}{2}\big)~,\end{equation}
valid for $y$ imaginary. Moreover, this combination is explicitly free of branch points due to the shift $1/2$.

\bibliographystyle{utphys}
\providecommand{\href}[2]{#2}\begingroup\raggedright\endgroup

\end{document}